\documentclass[11pt,a4paper]{article}
\pdfoutput=1
\usepackage{jinstpub}
\usepackage{graphicx}
\usepackage{subfigure}
\usepackage{lineno}
\modulolinenumbers[1]
\usepackage{hyperref} 
\usepackage{tabularx,booktabs}
\usepackage{multirow}
\usepackage[section]{placeins}

\def\nin{\noindent}

\usepackage{xspace}
\usepackage{color}
\usepackage{xcolor}
\usepackage{rotating}
\definecolor{light-gray}{gray}{0.8}

\newcommand{\GeVc}{\ensuremath{{\rm GeV/}c}\xspace}

\newcommand{\MeV}{\ensuremath{\rm MeV}\xspace}
\newcommand{\GeV}{\ensuremath{\rm GeV}\xspace}

\newcommand{\cms}{\ensuremath{{\rm cm}/s}\xspace}

\newcommand{\mum}{\ensuremath{\mu{\rm m}}\xspace}

\newcommand{\average}[1]{\ensuremath{\langle #1 \rangle}\xspace}

\newcommand{\cmnt}[1]{}

\title{Picosecond Timing Resolution Measurements of Low Gain Avalanche Detectors with a 120 GeV Proton Beam for the TOPSiDE Detector Concept}

\author[a,1]{M. Jadhav,\note{Corresponding author.}}
\author[a]{W. Armstrong,}
\author[a]{I. Cloet,}
\author[a]{S. Joosten,}
\author[b]{S. M. Mazza,}
\author[a]{J. Metcalfe,}
\author[a]{Z.-E. Meziani,}
\author[b]{H.F.-W. Sadrozinski,}
\author[b]{B. Schumm,}
\author[b]{and A. Seiden}

\affiliation[a]{Argonne National Laboratory,\\9700 S Cass Ave, Lemont, IL 60439, U.S.A.}
\affiliation[b]{SCIPP, Univ. of California Santa Cruz,\\1156 High Street, Santa Cruz, CA 95064, U.S.A.}

\emailAdd{mjadhav@anl.gov}

\abstract{This paper presents results that take a critical step toward proving 10 ps timing resolution's feasibility for particle identification in the TOPSiDE detector concept for the Electron-Ion Collider.
Measurements of LGADs with a thickness of 35 \mum and 50 \mum are evaluated with a 120 \GeV proton beam. The performance of the gain and timing response is assessed, including the dependence on the reverse bias voltage and operating temperature. The best timing resolution of UFSDs in a test beam to date is achieved using three combined planes of 35 \mum thick LGADs at -30~$^\circ$C with a precision of 14.3 $\pm$ 1.5 ps.}

\keywords{Timing detectors; Particle tracking detectors}

\arxivnumber{2010.02499} 

\begin{document}
\maketitle
\flushbottom

\section{Introduction}\label{sec:introductionsilicondetector}
\nin Ultra-Fast Silicon Detectors (UFSD)~\cite{Sadrozinski_2017} are a novel type of silicon detector that simultaneously provides spatial and timing resolution. One kind of UFSD, the Low-Gain Avalanche Detector (LGAD)~\cite{PELLEGRINI201412}, relies on an internal charge multiplication mechanism, or an avalanche effect, that is introduced in a controlled manner by the implantation of an appropriate acceptor or donor dopant layer. The internal charge multiplication process provides a moderate internal gain (up to 50) that increases the detector signal output, which further increases the signal-to-noise (S/N) ratio~\cite{Cartiglia:2016voy}.

UFSDs are targeted at a range of new opportunities from space science, mass spectroscopy, medical science to nuclear and particle physics~\cite{SADROZINSKI2013226}. The precision timing measurement combined with the high granularity spatial measurement enables a 4-Dimensional detector concept for particle detectors. A precise timing response translates into excellent time-of-flight (ToF) measurements, thus providing good particle identification (PID). The Timing Optimized PID Silicon Detector (TOPSiDE) concept~\cite{Repond:2018K9, Armstrong:2018, Joosten:2020} for the Electron-Ion Collider (EIC) at the Brookhaven National Laboratory is one such application that implements a 4D concept requiring precision timing on the order of 10 ps. With such precision, pion/kaon separation up to 7 \GeVc is expected using the ToF method with the efficiency of 90$\%$~\citep{Repond:2019hth}. It provides the precision measurements required for the EIC's broad physics program; some major topics include deep inelastic structure functions of protons and nuclei, spin structure functions, generalized parton distributions, and transverse momentum dependent distributions~\cite{Accardi:2012qut}. Here, the results demonstrate a 15 ps timing resolution with three layers of 35 $\mu$m thick sensors. It is expected that with thinner sensors, the 10 ps goal is readily achievable. 

\FloatBarrier

\section{Low Gain Avalanche Detectors}\label{sec:lgad}

\begin{figure}[ht]
\centering
\includegraphics[width=0.94\linewidth]{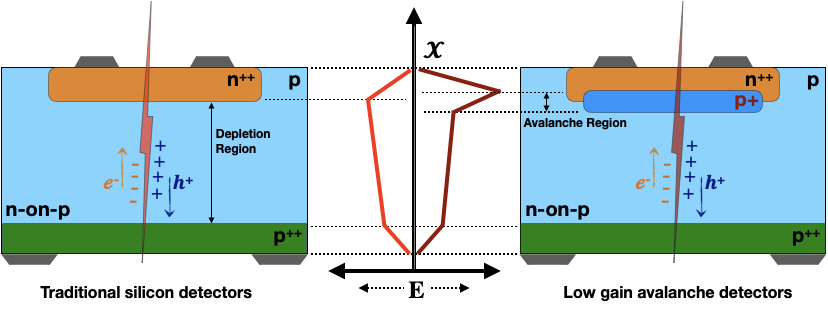}
\caption[The LGAD versus traditional silicon detector.]{Traditional silicon detector (left) and low-gain avalanche detector (right).}	
\label{fig:lgadcompare}
\end{figure}

\nin LGADs are distinguished from traditional silicon sensors by the gain layer beneath the electrode contacts, as shown in figure~\ref{fig:lgadcompare}.
The charge carrier collection time inside the silicon is limited by saturation of electron drift velocity at about 10$^7$ \cms and takes $\sim$1 ns to collect the charge for a $\sim$100 $\mu$m thick sensor~\cite{Sadrozinski_2017, SADROZINSKI20147}. The thicker sensors are required to keep the S/N to a level acceptable to the electronics. An LGAD relies on a gain layer to increase the signal by a factor of 10 to 100, allowing thinner sensors and a shorter drift distance with shorter collection time~\cite{Sadrozinski_2017, SADROZINSKI20147}. 

The first LGAD was fabricated by Centro Nacional de Microelectr\'{o}nica (CNM-IMB), Barcelona, Spain, with a thickness of 300 \mum and maximum gain of 10 at a bias voltage of 300 V at -10 $^\circ$C~\cite{PELLEGRINI201412}. A timing resolution of the order of 100 ps is reported for these sensors~\cite{Cartiglia:2016, Arcidiacono:2016}. Since then, several LGADs with much thinner thickness and improved gain have been studied~\cite{Cartiglia:2016, SADROZINSKI201618}. The best timing precision recorded so far with a single UFSD under test is 18 ps for LGAD with a thickness of 50 \mum and the gain of $\sim$70 at -20 $^\circ$C~\cite{ZHAO2019387}. S. Mazza, et al. ~\cite{Mazza_2020} also reported the timing resolution of 17 $\pm$ 1 ps for the trigger sensor. The best timing precision measured in a test beam scenario is 27 ps with a single LGAD sensor with thickness of 45 \mum (gain 70)~\cite{Cartiglia:2016voy} and 50 \mum (gain 27)~\cite{MINAFRA201788} at 180 GeV pion beam at CERN. The best timing reported with three 45 \mum LGADs together in a test beam set-up is 16 ps~\cite{Cartiglia:2016voy}.

\FloatBarrier

\section{Experimental setup}\label{sec:experimentalsetup}

\nin Measurements of several types of LGADs are performed with a 120 GeV proton beam at the Fermilab Test Beam Facility (FTBF). The trigger sensor's timing resolution is measured using the $^{90}$Sr $\beta$-telescope setup in the laboratory. The UFSD devices are characterized for charge collection efficiency, gain, and timing resolution for different bias voltages and temperatures. The following subsections describe the samples and test setup. 

\subsection{LGAD Devices and Electrical properties}\label{subsec:lgaddevice}

\nin The LGADs with different thicknesses, doping concentrations, and sensor types (pad, pixels, strips, AC-LGADs) have been measured for timing resolution and radiation hardness. The LGADs tested and presented in this paper are manufactured by Hamamatsu Photonics K.K. (HPK) and tagged as HPK-1.2 and HPK-3.1. The sensors are $n$-on-$p$ type with a thickness of 35 \mum and 50 \mum, respectively, whereas the physical thickness, including substrate wafer, varies between 300 to 350 \mum. HPK-1.2 and HPK-3.1 both have an active pad area of 1.3 $\times$ 1.3 mm$^2$. Both the LGADs are doped with Boron to create an internal multiplication layer. HPK-1.2 has shallow implantation, and higher resistivity with a breakdown voltage of around 270 V at room temperature. The HPK-3.1 has deeper implantation with a breakdown voltage slightly above 245 V at room temperature. The capacitance of the LGADs vary with the total active bulk and doping profile. It is measured to be 5.35 pF and 3.9 pF for LGADs HPK-1.2 and HPK-3.1, respectively. The detailed description of the LGADs can be found in references \cite{Seiden:2020inw,Padilla:2020sau,Cindro:2020qxe,Yang:2020qis}.
%
\begin{table}[ht]
  \centering
  \caption[LGAD device descriptions.]{The description of LGAD devices under test.}
  \label{tab:lgaddevice}
  \begin{tabular*}{\textwidth}{@{\extracolsep{\fill}} c|c|c|c|c|c|c}
    \toprule
    \multicolumn{1}{c}{Sensors} & \multirow{2}{*}{Type} & \multicolumn{1}{c}{Thickness} & \multicolumn{1}{c}{Pad Area} & \multicolumn{1}{c}{C} & \multicolumn{1}{c}{Rise Time} & \multicolumn{1}{c}{Breakdown} \\
 					HPK	& &(\mum) & (mm$^2$) & (pF) & (10-90$\%$) & Voltage\\
    \midrule
     \multirow{1}{*}{3.1} & $n$-on-$p$	& \multirow{1}{*}{50} 	& \multirow{1}{*}{1.3 $\times$ 1.3} 	& \multirow{1}{*}{3.9} & 470 ps & 245 V \\[2ex]
     \multirow{1}{*}{1.2} & $n$-on-$p$	& \multirow{1}{*}{35} 	& \multirow{1}{*}{1.3 $\times$ 1.3} 	& \multirow{1}{*}{5.35} & 375 ps & 270 V \\[2ex]
    \bottomrule
  \end{tabular*}
\end{table}  

Additionally, an HPK-8664, a $p$-on-$n$ type LGAD sensor, is evaluated. The HPK-8664 has a round pad with approximately 1 mm$^2$ of a circular active area without any guard ring protection. With the breakdown voltage of 430 V, this sensor is used as a trigger during data taking. 

\subsection{Sample Setup}\label{subsec:samplesetup}
\begin{figure}[ht]
\centering
\subfigure[Read-out board.]{
\label{fig:readoutboardfull}
\includegraphics[width=0.44\textwidth]{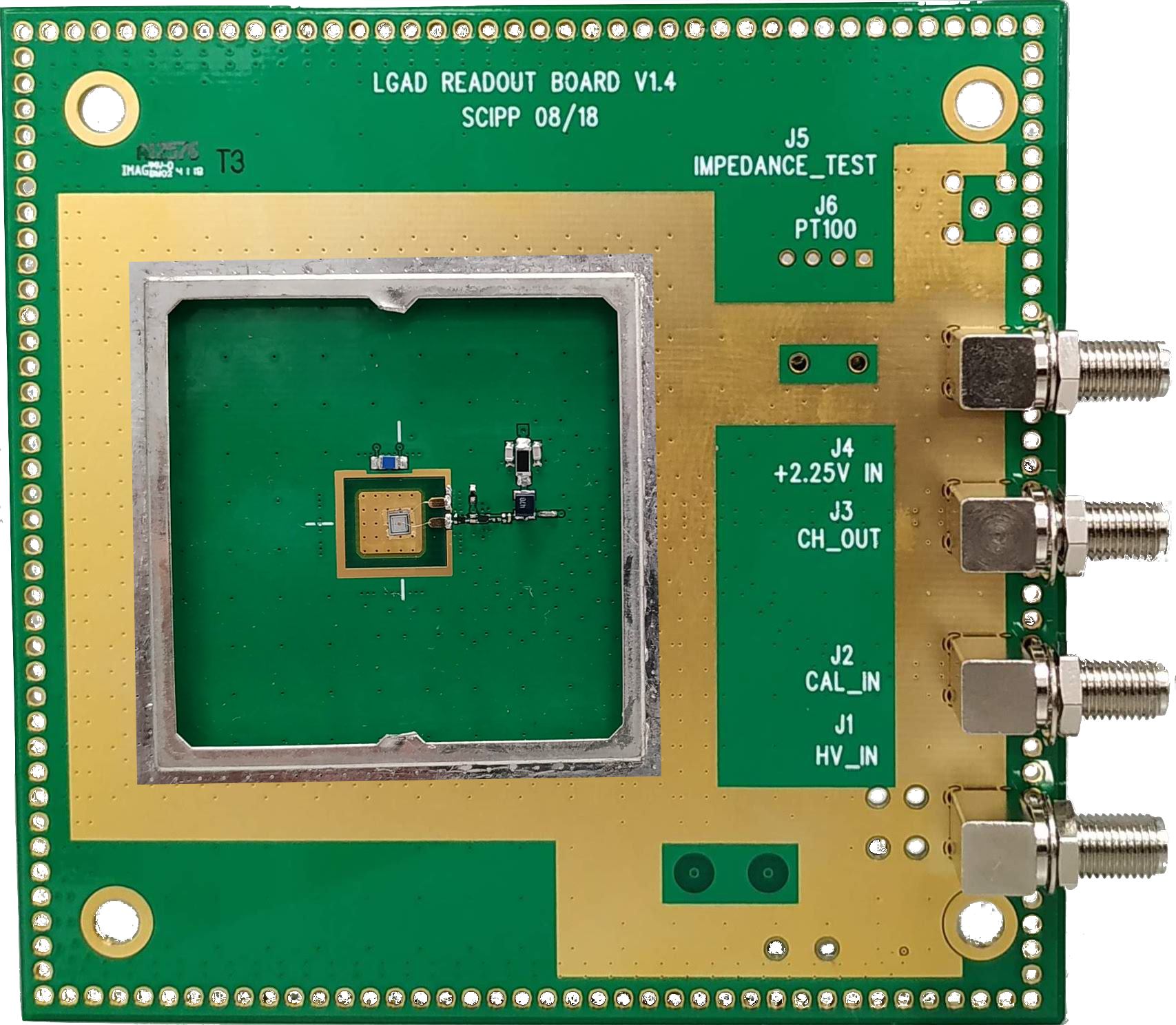}}
\subfigure[UFSD alignment set-up.]{
\label{fig:setup3d}
\includegraphics[width=0.49\textwidth]{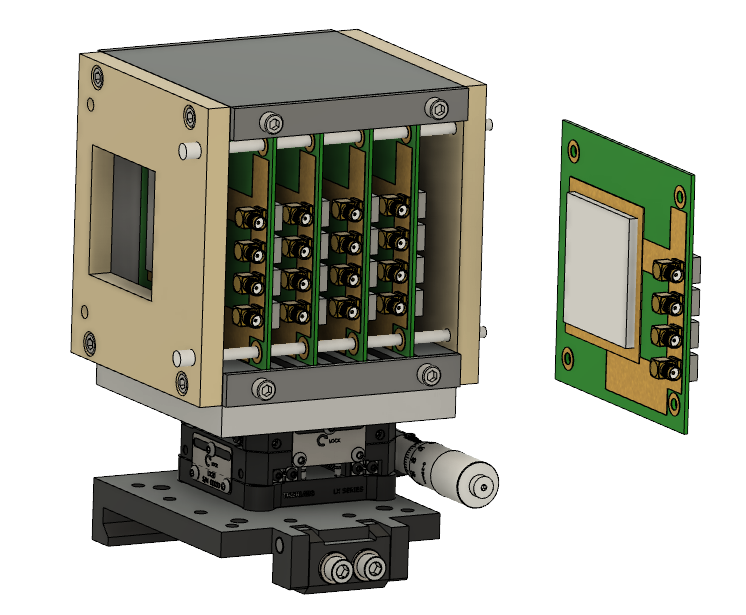}}
\caption[The alignment frame and read-out board design.]{A read-out front-end board with discrete components to collect the signal (left) and the alignment frame (right).}
\label{fig:measurementsetup3d}
\end{figure}

\nin The LGAD sensors were mounted on a $\sim 10 \times 10$ cm$^2$ read-out board, shown in figure~\ref{fig:readoutboardfull}. The single-channel read-out board contains wide bandwidth ($\sim$2 GHz) and low noise inverting amplifier with a gain of 10 and has been used in previous studies\cite{Sadrozinski_2017, Cartiglia:2016voy}. The schematic of the read-out circuit is provided in A. Seiden et al.~\cite{Seiden:2020inw}. The inverting amplifier uses a high-speed SiGe Bipolar transistor (Infineon-BFR840L3RHESD) with a trans-impedance of about 470 $\Omega$. The inverting amplifier is followed by a second stage 20dB external amplifier (a mini-circuit TB-409-52+) with a gain of 10. The combined trans-impedance of the read-out is about 4700 $\Omega$. The value of the total trans-impedance is simulated using LTspice simulation and examined using lab experimental tests (details are provided in Z. Galloway et al.~\cite{GALLOWAY201919}). The current drawn through the amplifier is around 15-17 mA. Considering this the power consumption of the amplifier is below 40 mW. 

\begin{figure}[ht]
\centering
\subfigure[On bench setup in the lab.]{
\label{fig:labsetup}
\includegraphics[width=0.475\textwidth]{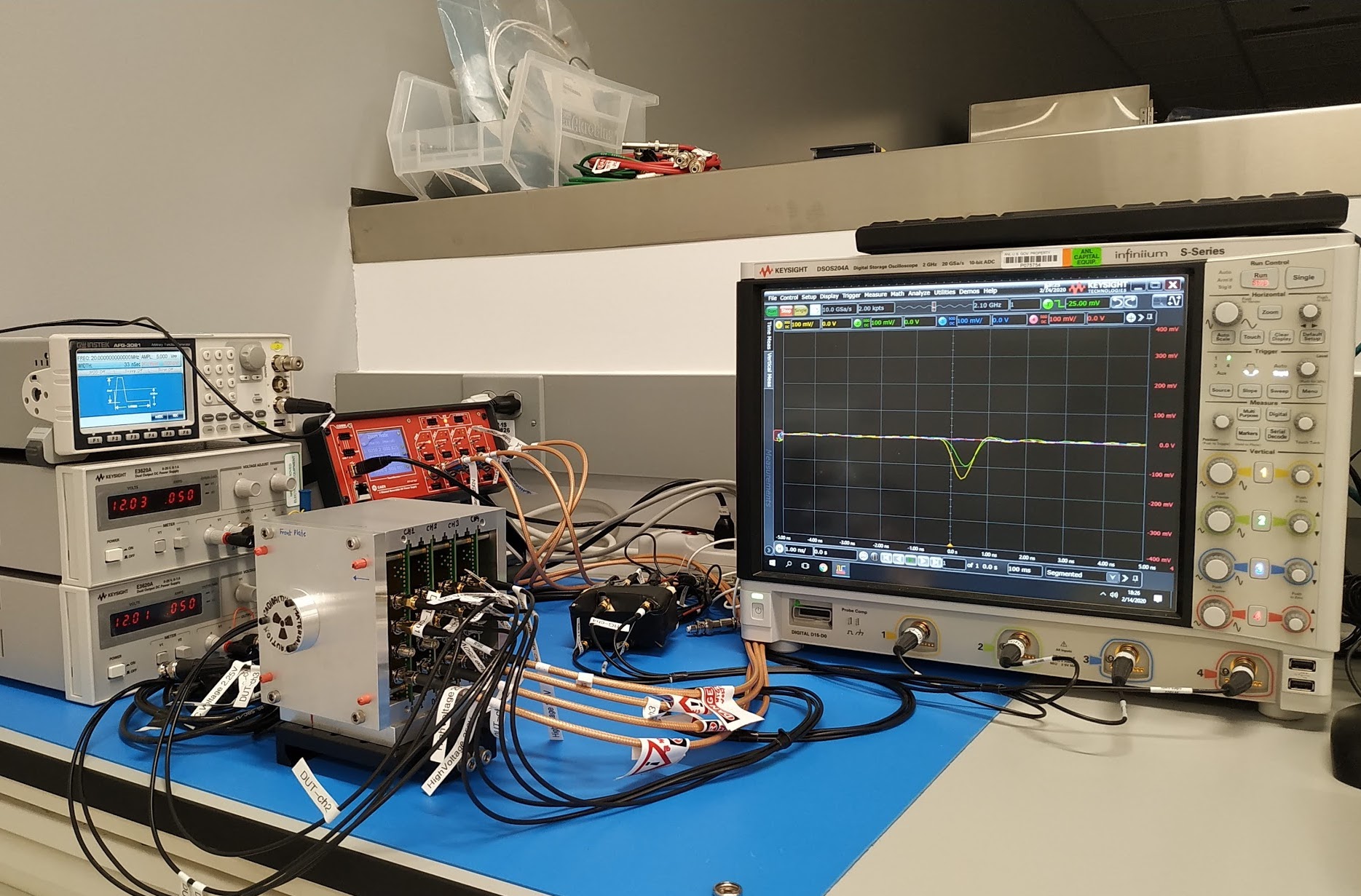}}
\subfigure[Test beam setup at Fermi Lab.]{
\label{fig:testbeamsetup}
\includegraphics[width=0.445\textwidth]{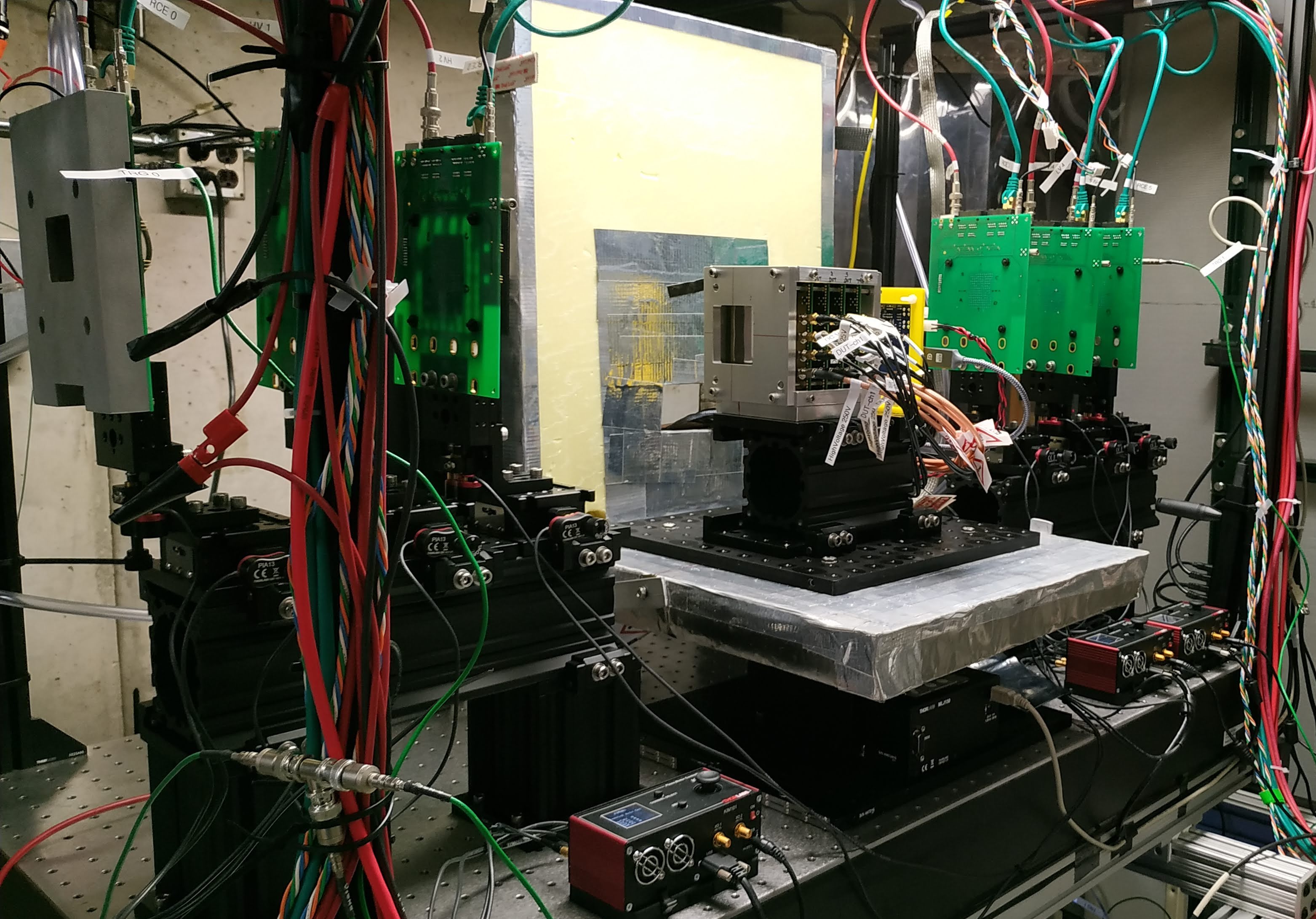}}
\caption[The setup at the lab and the test beam.]{Test setup for timing measurements of LGADs using time-of-flight (TOF) technique for minimum ionizing particles (MIP) using Sr$^{90}$ source in the lab (left) and at the Fermi lab test beam facility (right).}
\label{fig:measurementsetup}
\end{figure}

The sensors are glued to the 5 mm square pad on the read-out board, and the guard ring is grounded. Two or more read-out boards are aligned back-to-back, as shown in the figure~\ref{fig:setup3d} using alignment rods passing through the holes at each corner of the alignment frame and read-out boards. The alignment box is designed to use up to four sample boards. It is used in both setups, the $^{90}$Sr $\beta$-Telescope and test beam measurement setup as shown in the figure \ref{fig:labsetup} and \ref{fig:testbeamsetup} respectively. In the $^{90}$Sr $\beta$-Telescope it has a source holder aligned with LGAD positions on the read-out board. Whereas, for test beam measurements, the Aluminum plates have an open window to pass the beam and prevent the absorption/energy loss of the incoming particles. The incident particle radiations pass perpendicularly through the sensors' active region.

In each case, one sensor is designated as the trigger sensor (TRG), and the other sensors are considered as the devices under test (DUTs). The samples are biased using CAEN DT1471ET high voltage supply. The amplified signal from the read-out is recorded using a 2.2 GHz Keysight DSOS204A digital oscilloscope. The sampling rate is 20 GSa/s with a time discretization of 50 ps with two active channels. In contrast, the sampling rate reduces to 10 GSa/s with a time discretization of 100 ps while using all four oscilloscope channels. The events are registered from all DUTs when the trigger sensor provides a signal above the threshold value. The oscilloscope captures the events as a signal pulse waveform. The raw data is acquired and stored in a computer using a Python-based PyVisa interface, referred to as the DAQ framework.

\subsection{\texorpdfstring{$^{90}$Sr $\beta$-Telescope Setup}{Sr90 beta-Telescope setup}}\label{subsubsec:betatelescope}

The $^{90}$Sr $\beta$-source measurements are performed by mounting the source in the alignment box. The $\beta$ particle with energy 0.546 \MeV can only penetrate two sensors before being absorbed. Thus, only two oscilloscope channels are used, allowing for a 20 GSa/s sampling rate. Cold measurements are done by placing the setup in an environmental chamber.

\subsection{Fermilab Test Beam Setup}\label{subsubsec:testbeamsetup}

\nin The test beam measurements are performed at the Fermilab Test Beam Facility (FTBF) using a 120 GeV proton beam~\cite{fermilabtestbeam:2020}. For the test beam campaign, four LGAD sensors are used simultaneously, utilizing all four oscilloscope channels. However, this limits the sampling rate to 10 GSa/s with a time discretization of 100 ps. The time resolution measurements are compared to those with a sampling rate of 20 GSa/s, and the results are consistent within the uncertainties. The trigger is the fourth LGAD sensor in the alignment box placed downstream in the beamline. The data is acquired at room temperature for two types of UFSDs as described above. Additionally, the low-temperature measurements are performed for the HPK-1.2 at -30 $^\circ$C. The temperature reported in this paper is the temperature of the air inside the cold enclosure where the measurement system is housed. It is maintained at a low temperature using a cold plate cooled by a Julabo FP89-ME recirculating chiller. The humidity inside the box is kept below the dew point by using Nitrogen purging.

\begin{figure}[ht]
\centering
\includegraphics[width=0.5\textwidth,clip,trim=2mm 0mm 18mm 10mm]{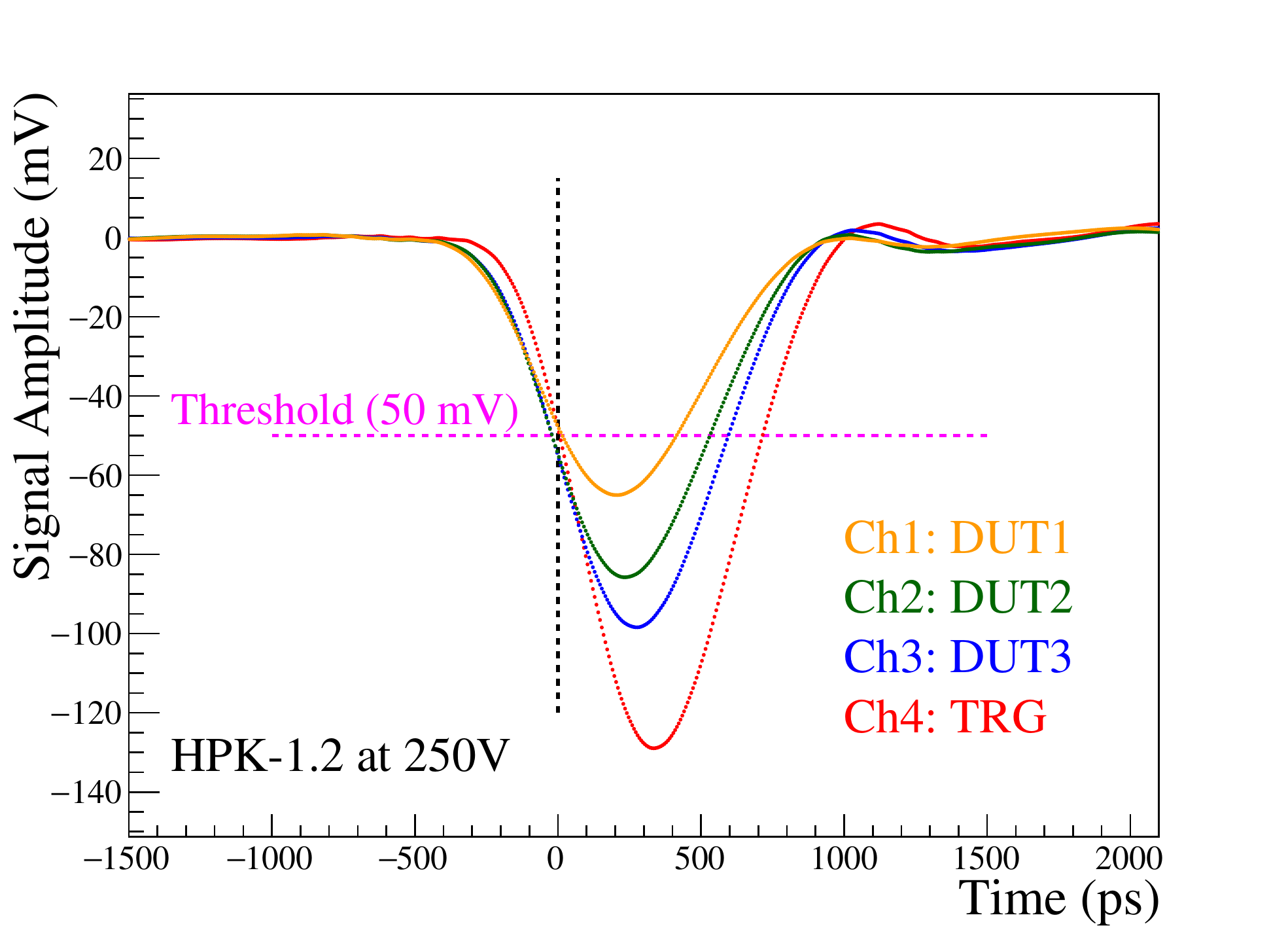}
\caption[The signal pulse from TRG and DUT.]{The event with a signal pulse from TRG and DUT. As detectors are aligned back-to-back, such events are considered to be from minimum ionizing particles (MIP).}
\label{fig:outputpulsesignal}
\end{figure}

The test beam data is collected in spills of 4 seconds per minute. The symmetric beam profile with $\sigma$-value of 2-4 mm is used, and intensity is maintained between 100K to 200K protons every spill. The beam is adjusted to get an instantaneous trigger rate between 1 and 10 Hz. All the DUTs in the setup are operated at the same bias voltage in each run. The data is acquired at biases ranging from 200 V to 255 V for UFSDs; HPK-1.2, and HPK-3.1 at different temperature conditions. The maximum bias voltage is defined by the breakdown conditions for a particular UFSD at a specific temperature. The trigger is supplied with a fixed bias voltage of 425 V throughout the test beam run. The leakage current through all the sensors was stable at around 15-25 nA during the test run. 

Figure~\ref{fig:outputpulsesignal} shows an example of one of the events at a test beam run where all four UFSDs are fired. In the event, all four HPK-1.2 UFSDs are operated at a bias voltage of 250 V providing pulse signal with a rise time around 350 ps as shown in section~\ref{subsec:risetime}.

\FloatBarrier

\section{Data Analysis Method}\label{sec:dataanalysis}

\begin{figure}[ht]
\centering
\subfigure[Calculation of CFD.]{
\label{fig:cfdcalculation}
\includegraphics[width=0.5\textwidth]{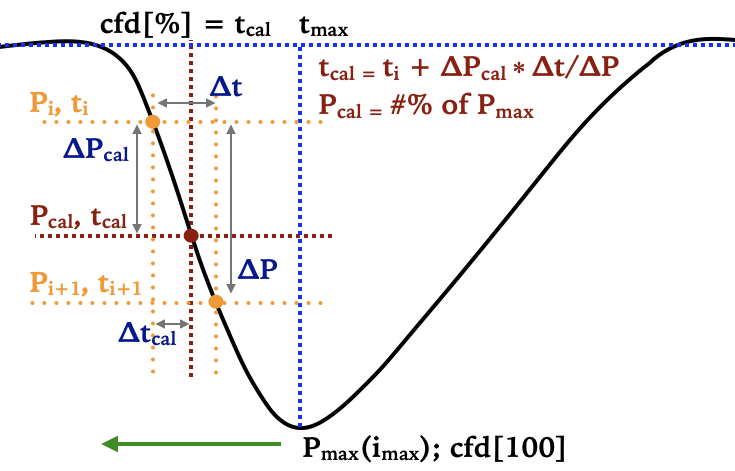}}
\subfigure[Signal pulses and CFD distribution.]{
\label{fig:cfddistribution}
\includegraphics[width=0.45\textwidth]{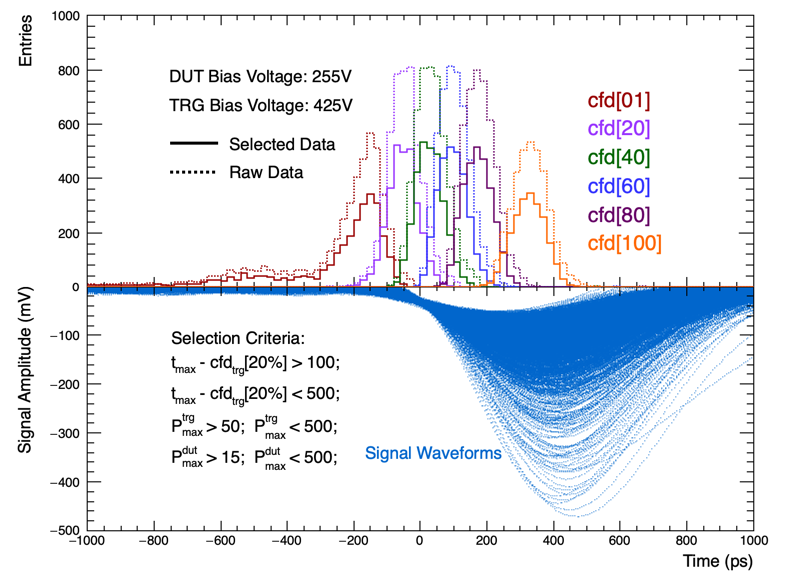}}
\caption[The CFD calculation and distribution.]{The CFD value calculation using the linear interpolation method provides the timing information at a certain fraction of maximum signal amplitude (left). The time distribution at different CFD values compared to signal waveform data for the HPK-8664 trigger detector at bias voltage 425V taken with 120 GeV Proton beam (right).}	
\label{fig:cfd}
\end{figure}

\nin The data analysis follows a similar procedure as~\cite{Cartiglia:2016voy, Mazza_2020}. It uses only those variables which would be available with a hypothetical read-out chip, like time-of-arrival (ToA), signal amplitude, and time-over-threshold (ToT). The time at which the signal pulse crosses a certain fraction of the maximum signal amplitude is considered time-of-arrival. The corresponding amplitude is referred to as the CFD value. For example, ${\rm cfd}[20]$ indicates the time with CFD value equals to 20$\%$ of maximum pulse amplitude. The information plays a critical role in calculating different parameters, like the time difference between DUT and trigger, rise time, jitter, etc. The method is called a constant fraction discriminator (CFD) method. The CFD method is highly effective in correcting the time-walk effect, where the oscilloscope digitization helps calculate time-of-arrival precisely by linear interpolation. Figure~\ref{fig:cfdcalculation} shows the calculation of CFD value and the corresponding time using a linear interpolation method. The distribution of time at different CFD values (fractions) and its comparison with signal waveform data is shown in figure~\ref{fig:cfddistribution}.

\begin{figure}[ht]
\centering
\includegraphics[width=0.5\textwidth,clip,trim=0mm 0mm 0mm 10mm]{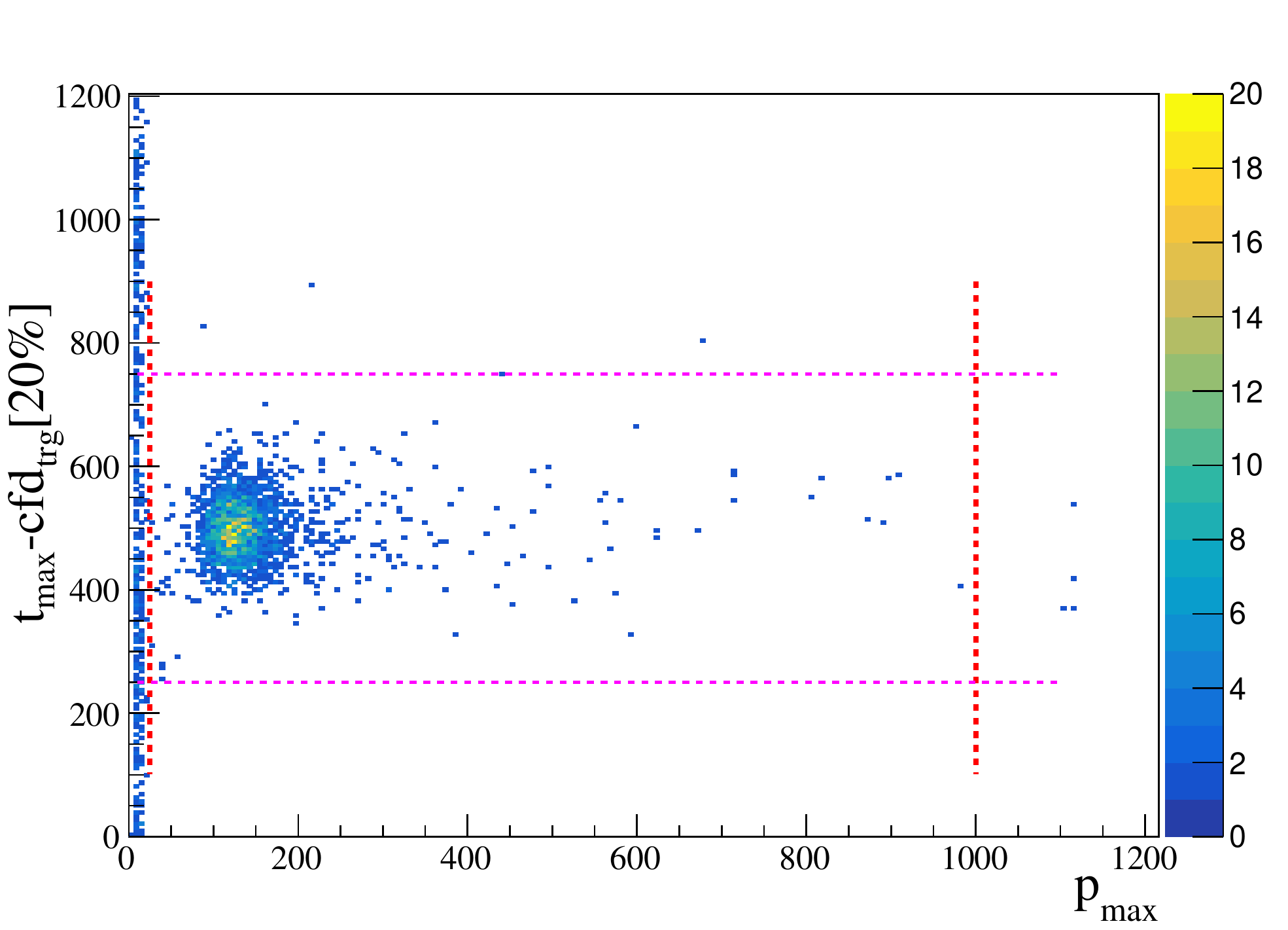}
\caption[Event selection to avoid saturated and non-gain events.]{Event selection based on the distribution of time difference versus signal amplitude for DUT HPK-1.2 operated at bias voltage 250V with 120 GeV proton beam at Fermilab test beam facility.}
\label{fig:cutscoincidence}
\end{figure}

The selection criteria applied to events with a valid trigger pulse are mainly based on signal amplitude and time of arrival. The maximum signal amplitude, ${\rm P}_{max}$ of all UFSDs, should be at least five times larger than the noise level (2.5-4.5 mV, section~\ref{subsec:noiseandjitter}), and the oscilloscope or the read-out chain should not saturate it. The second selection is on the time difference between the DUT and the trigger.  The time difference criterion reduces the contribution from the non-gain events or noise. As shown in the upper panel of figure~\ref{fig:cfddistribution}, the selection rules allow removing the low-gain tail effect on the distribution's left side. Figure~\ref{fig:cutscoincidence} shows the amplitude selection along the horizontal axis and time selection along the vertical axis. The detection efficiency for the LGAD is proven to be almost 100$\%$ for moderate gain~\cite{CERN-LHCC-2020-007}. In this case for the collected data, the selection efficiency varies between 25$\%$ to 50$\%$. 

\nin The data is analyzed to study the signal amplitude distribution, collected charge, gain, jitter, rise time, and timing resolution. The following sub-sections provide analysis details of these variables.

\subsection{Time Resolution}\label{subsec:timeresolution}
\nin The time resolution, $\sigma_{t}$ of the detector system can be expressed as a sum of different contributions \cite{Sadrozinski_2017, CARTIGLIA2015141},  
\begin{equation}\label{eq:timeresolution}
\sigma_{t}^{2} = \sigma_{Jitter}^{2} + \sigma_{LandauNoise}^{2} + \sigma_{TimeWalk}^{2} + \sigma_{Distortion}^{2} + \sigma_{TDC}^{2}.
\end{equation}

The predominant contribution to the timing resolution is from jitter and Landau noise. Jitter is the time uncertainty caused by the early or delayed firing of a comparator due to noise in the signal itself or the electronics. The CFD method reduces the time-walk contribution and is considered negligible. Likewise, the signal distortion is negligible in silicon for the saturated drift velocity and uniform weighting field. The uniform weighting field is achievable using ``parallel plate" geometry with a large active area compared to sensor thickness~\cite{Sadrozinski_2017}. The contribution from the time-to-digital converter (TDC) is equal to the timing uncertainty, $\Delta$T/$\sqrt{12}$, where $\Delta$T is the least significant bit of the TDC. The TDC effect is minimal in most of the cases and ignored in this paper. 

\subsubsection{Jitter and Landau Fluctuation}\label{subsubsec:jitterlandaunoise}
\nin Jitter is directly proportional to the noise and minimized by having a higher slew-rate and low intrinsic noise. 
\begin{equation}\label{eq:jitter}
\sigma_{jitter} = \frac{Noise}{{\rm d}V/{\rm d}t}.    
\end{equation}

Wherein, the noise is determined as the RMS fluctuation of the oscilloscope's baseline trace and calculated using $1/4^{th}$ of total points in the waveform from the beginning of the pulse.

Another contribution to timing performance is Landau noise, which is introduced by a particle's non-uniform charge deposition along its passage. The Landau noise is insensitive to the gain value and found to be dependent on CFD settings. The Landau noise decreases with the thickness of the sensor~\cite{Sadrozinski_2017, GALLOWAY201919}. It has been observed that jitter and Landau noise contribute almost equally and significantly in time resolution for the LGAD sensors~\cite{Sadrozinski_2017}.

\subsubsection{Timing Response}\label{subsubsec:timingresponse}

\nin The timing response of the LGAD sensor is measured using the time difference between signals from DUT and the trigger sensor. The Gaussian function is used to fit the time difference ($\Delta$t) distribution, which attributes the contributions from the equation~\ref{eq:timeresolution}. The fitting parameter; $\sigma_{(DUT-TRG)}$ provides the timing resolution of the device under test using a quadrature sum:
\begin{equation}\label{eq:timeresolutionsigma}
\sigma_{DUT}^{2} + \sigma_{TRG}^{2} = \sigma_{(DUT-TRG)}^{2}.
\end{equation}

 The DUT and the trigger's timing resolution is denoted as $\sigma_{DUT}$ and $\sigma_{TRG}$ respectively. For a unique setup where DUT and TRG are the same type of LGAD, the equation is simplified as,
\begin{equation}\label{eq:timeresolutionsamesensor}
\sigma_{DUT} = \sigma_{(TRG-DUT)}/\sqrt{2}. 
\end{equation}

\subsection{Charge Collected and Gain}\label{subsec:gain}
\nin The timing resolution of the LGAD sensor also depends on the charge collection and gain. The energy deposited by the minimum ionizing particles in the active bulk of the LGAD sensor follows the Landau distribution before getting amplified by the internal gain process. 

The ratio of total charge after multiplication and the initial number of charge carriers gives the internal gain. The collected charge is calculated by dividing the pulse area by the input trans-impedance of the detector system. The pulse area is the integration of pulse waveform from 1 ns before the start of the pulse (i.e., zero-crossing) to 3 ns after the pulse reaches its maximum amplitude (i.e., t$_{max}$). The Weight-Field2 simulation calculates the initial charge collection using an identical PIN sensor without a gain layer.

\FloatBarrier

\section{Laboratory Results}\label{sec:laboratoryresults}

\begin{figure}[ht]
\centering
\subfigure[HPK-1.2 at 250 V.]{
\label{fig:HPK1dot2bias250}
\includegraphics[width=0.47\textwidth]{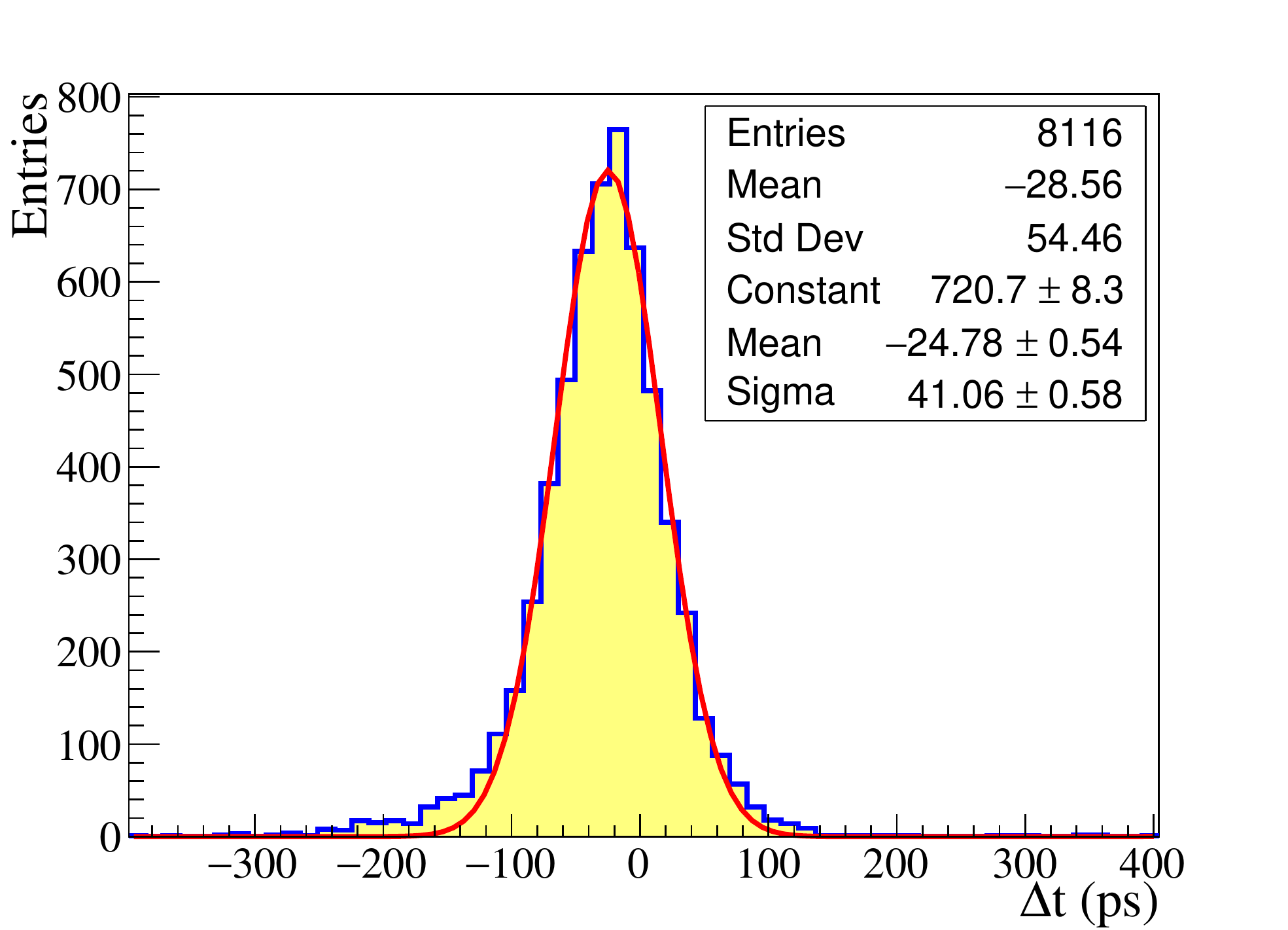}}
\subfigure[HPK-8664 at 425 V.]{
\label{fig:HPK8664bias425}
\includegraphics[width=0.47\textwidth]{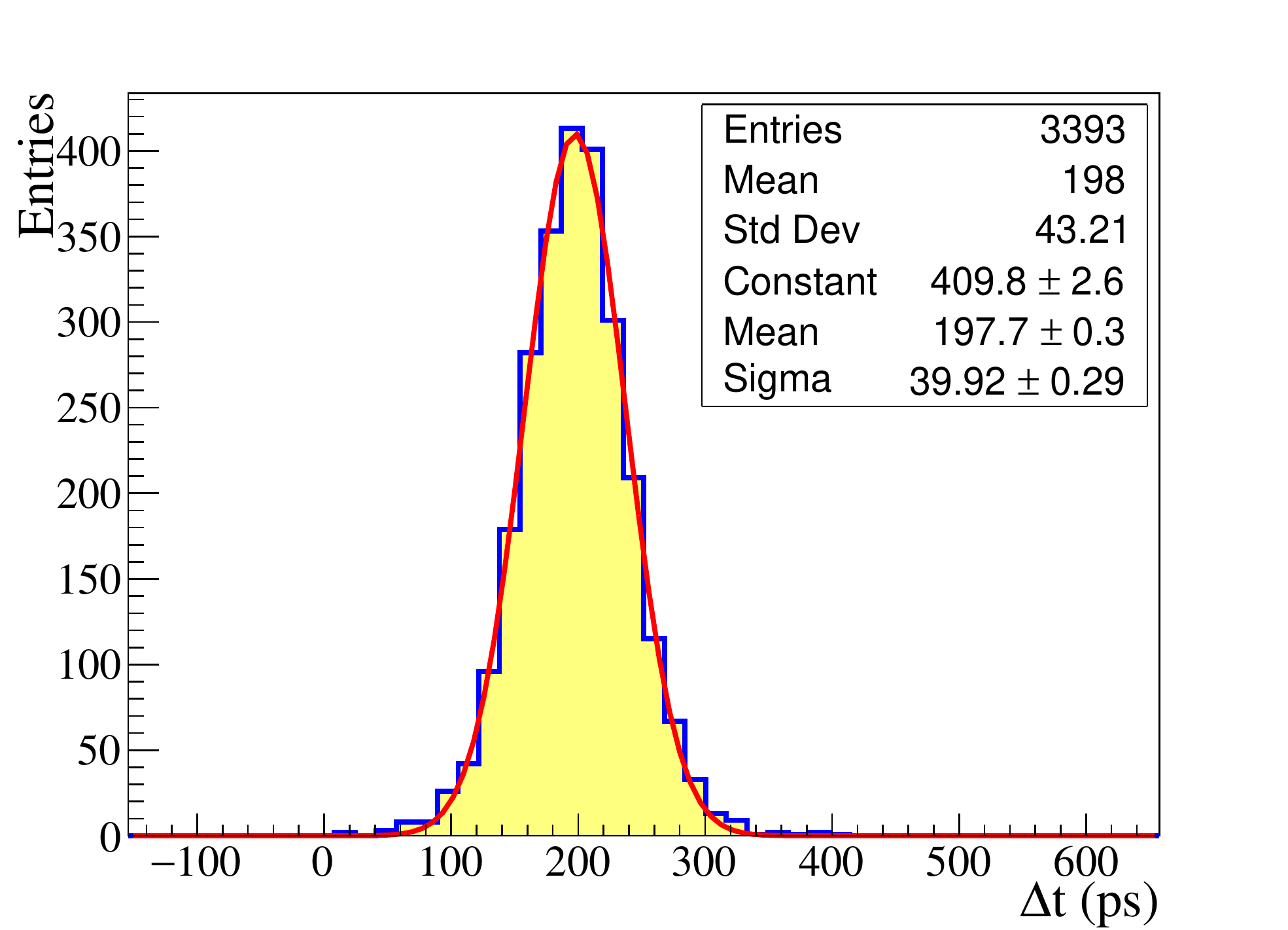}}
\caption[TOF between DUT and TRG planes.]{The distribution of time difference between the DUT and the trigger. The DUT and the trigger both are HPK-1.2 operated at 250 V (left). The time difference between HPK-8664 at 425 V and HPK-1.2 at 250 V (right).}
\label{fig:timingdistribution}
\end{figure}

\nin The timing resolution is first measured in the laboratory using the $\beta$ telescope as reference before taking measurements at the Fermilab Test Beam Facility (FTBF). The timing resolution of HPK-1.2 is measured using two identical LGADs. Figure~\ref{fig:HPK1dot2bias250} shows the timing difference at 50$\%$ CFD fractions, and the timing resolution is calculated using equation~\ref{eq:timeresolutionsamesensor}. The HPK-1.2 provides a timing resolution of 29.0 $\pm$ 0.4 ps for bias voltage 250 V at room temperature and 26.3 $\pm$ 0.9 ps for 210 V at -30 $^\circ$C. The timing resolution of HPK-8664 is then measured using HPK-1.2 as a trigger.  The HPK-8664 is operated at a bias voltage of 425 V and HPK-1.2 at 250 V. For the set-up with two different types of UFSDs, the time difference is taken at two different but optimized values of CFD fractions, i.e., 20$\%$ CFD for HPK-8664 and 50$\%$ CFD for HPK-1.2 sensor. The timing resolutions of HPK-8664 is calculated using equation~\ref{eq:timeresolutionsigma}. It is 27.4 $\pm$ 0.6 ps for bias voltage 425 V at room temperature. Whereas, at temperature -30 $^{\circ}$C, it is 24.2 $\pm$ 0.7 ps when operated at 390 V. The results are tabulated in the table~\ref{tab:timeresolutionlabtest}.  

\begin{table}[!ht]
  \centering
  \caption[Timing resolution for set of DUTs in the lab]{The timing resolution for HPK-1.2 and HPK-8664 measured in the lab using $^{90}$Sr $\beta$-source.}
  \label{tab:timeresolutionlabtest}
  \begin{tabular*}{\textwidth}{@{\extracolsep{\fill}} cccccc}
    \toprule
     \multirow{2}{*}{Temp. ($^\circ$C)}	& \multirow{2}{*}{DUT} & \multirow{2}{*}{TRG} & \multicolumn{2}{c}{Bias Voltage (V)} & \multirow{2}{*}{Timing Resolution (ps)}\\
  							&  &  & DUT & TRG & \\
  							    \midrule
    \multirow{2}{*}{25} 		       	& \multirow{1}{*}{HPK-1.2} 	& \multirow{1}{*}{HPK-1.2} 	& \multirow{1}{*}{250} & \multirow{1}{*}{250} & \multirow{1}{*}{29.0 $\pm$ 0.4}\\
    \cmidrule(lr){2-6}
                 		       	& \multirow{1}{*}{HPK-8664} 	& \multirow{1}{*}{HPK-1.2} 	& \multirow{1}{*}{425} & \multirow{1}{*}{250} & \multirow{1}{*}{27.4 $\pm$ 0.6}\\
                 		       	\midrule
    \multirow{2}{*}{-30} 		       	& \multirow{1}{*}{HPK-1.2} 	& \multirow{1}{*}{HPK-1.2} 	& \multirow{1}{*}{210} & \multirow{1}{*}{210} & \multirow{1}{*}{26.3 $\pm$ 0.9}\\
        \cmidrule(lr){2-6}
                 		       	& \multirow{1}{*}{HPK-8664} 	& \multirow{1}{*}{HPK-1.2} 	& \multirow{1}{*}{390} & \multirow{1}{*}{210} & \multirow{1}{*}{24.2 $\pm$ 0.7}\\

    \bottomrule
  \end{tabular*}
\end{table}  
\FloatBarrier

\section{Test Beam Results}\label{sec:testbeamresults}
\nin The following sections provide results from the Fermilab Test Beam Facility taking measurements with a 120 GeV proton beam at room temperature and -30 $^\circ$C. Charge collection, gain, signal amplitude, rise time, noise, jitter, and timing resolution are reported for HPK-1.2 and HPK-3.1 LGADs.  

\subsection{Charge Collection and Gain}\label{subsec:chargecollectionandgain}

\begin{figure}[ht]
\centering
\includegraphics[width=0.5\textwidth,clip,trim=0mm 0mm 25mm 25mm]{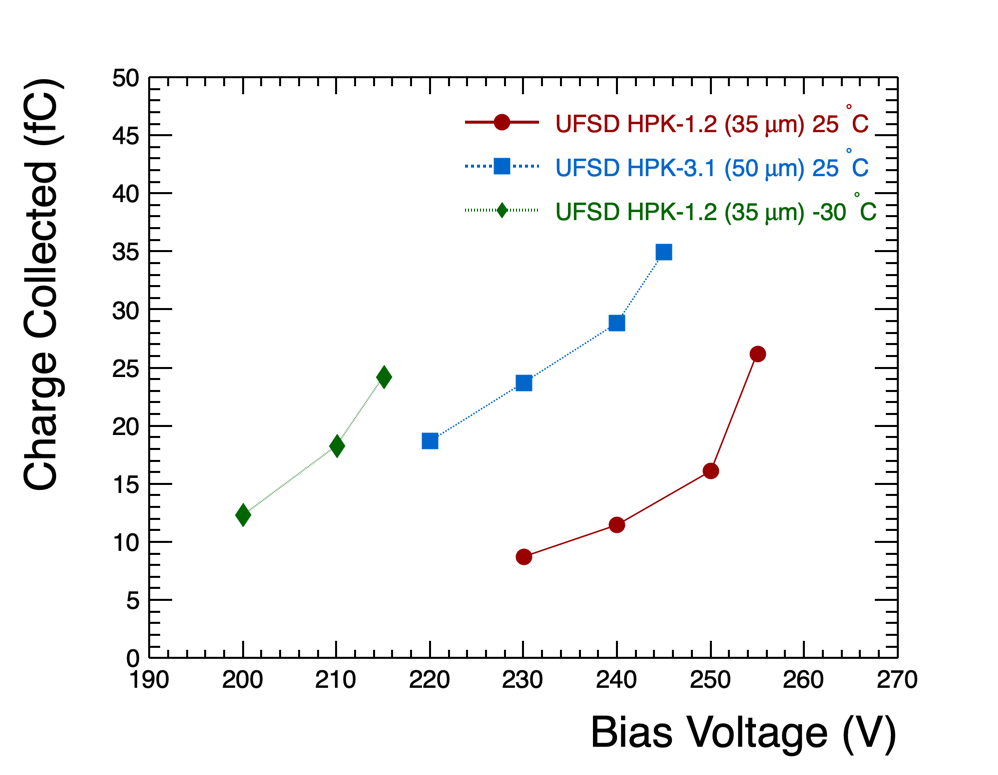}
\includegraphics[width=0.5\textwidth,clip,trim=0mm 0mm 25mm 25mm]{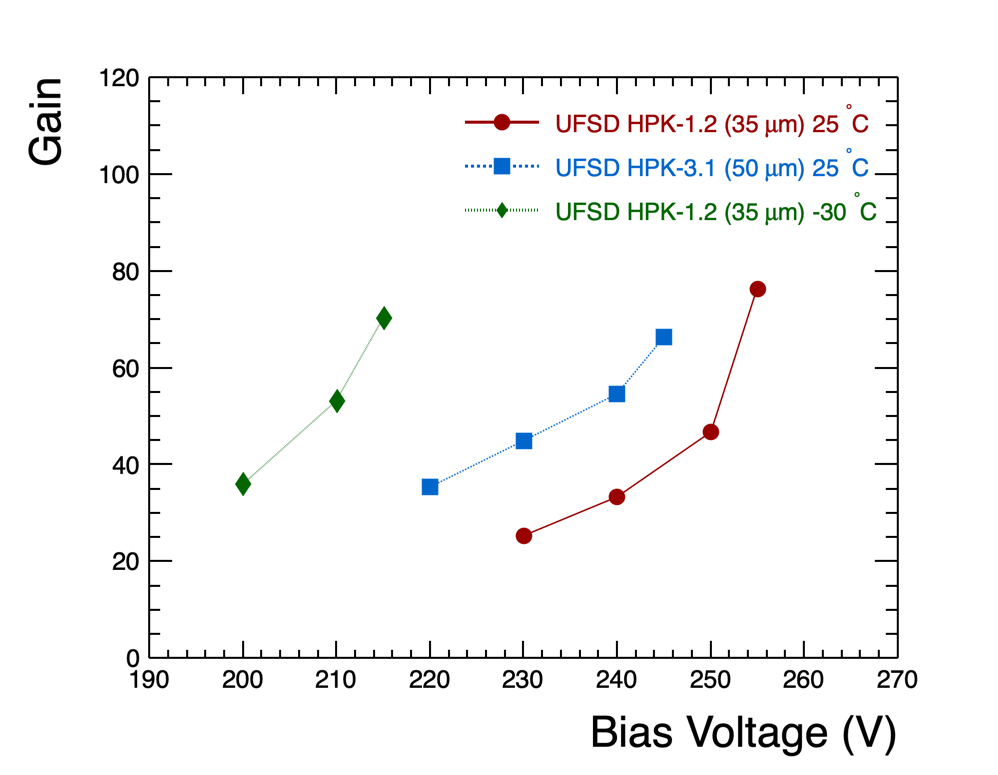}
\caption[Charge Collected vs. Bias Voltage.]{The charge collected (left) and the gain (right) as a function of bias voltage.}	
\label{fig:chargecollectedandgain}
\end{figure}

\nin The charge collection performance of HPK-1.2 and HPK-3.1 as a function of the bias voltage is shown in the left panel of figure~\ref{fig:chargecollectedandgain}. The figure shows the results taken at 25 $^\circ$C (room temperature) and -30 $^\circ$C. The rising trend of charge collection indicates the increased rate of charge multiplication along with bias voltage. By extrapolating results at room temperature, the figure also demonstrates that the charge multiplication rate at a certain bias voltage increases with reduced temperature but is restricted by reduced breakdown voltage at low temperatures. The right panel of figure~\ref{fig:chargecollectedandgain} shows the gain as a function of bias voltage for HPK-1.2 and HPK-3.1 at room temperature and -30 $^\circ$C. The simulated initial charge collection is 0.344 fC and 0.528 fC for 35 \mum and 50 \mum thick identical PIN sensors, respectively. The results show the exponential dependence of internal gain on the electric field and subsequently on the reverse bias voltage as expected in~\cite{PhysRev.109.1537, Sadrozinski_2017}. At room temperature, the maximum gain of 76 is achieved from HPK-1.2 biased at 255 V.

\subsection{Signal Amplitude}\label{subsec:signalamplitude}

\begin{figure}[ht]
\centering
\includegraphics[width=0.5\textwidth,clip,trim=5mm 5mm 25mm 23mm]{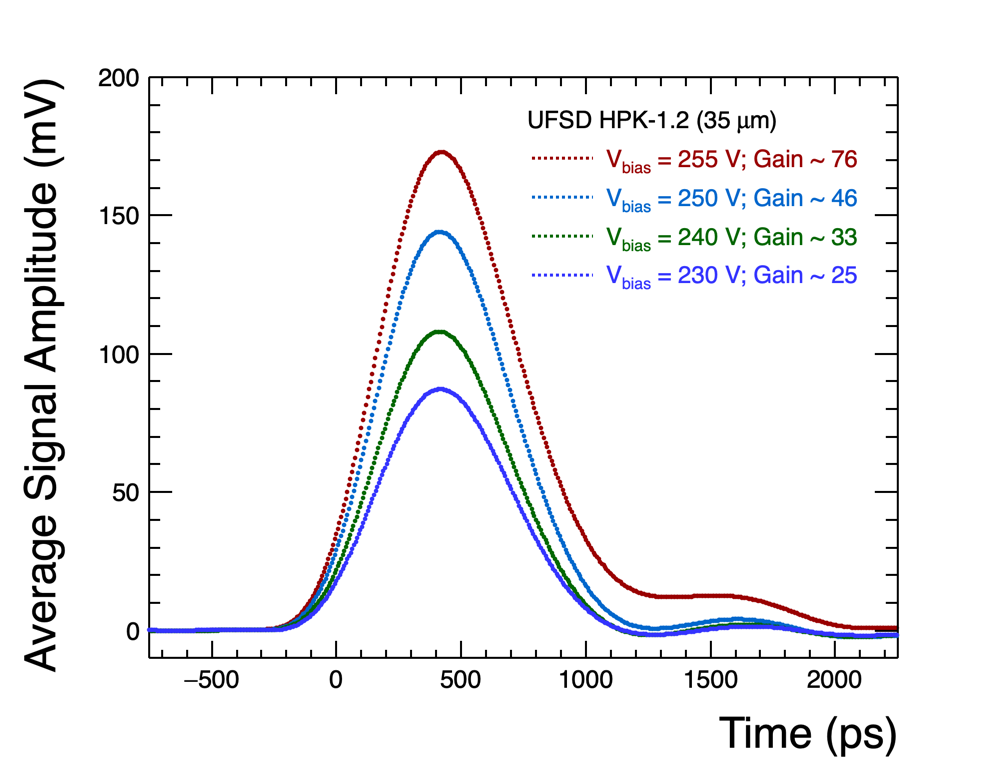}
\includegraphics[width=0.5\textwidth,clip,trim=5mm 5mm 25mm 23mm]{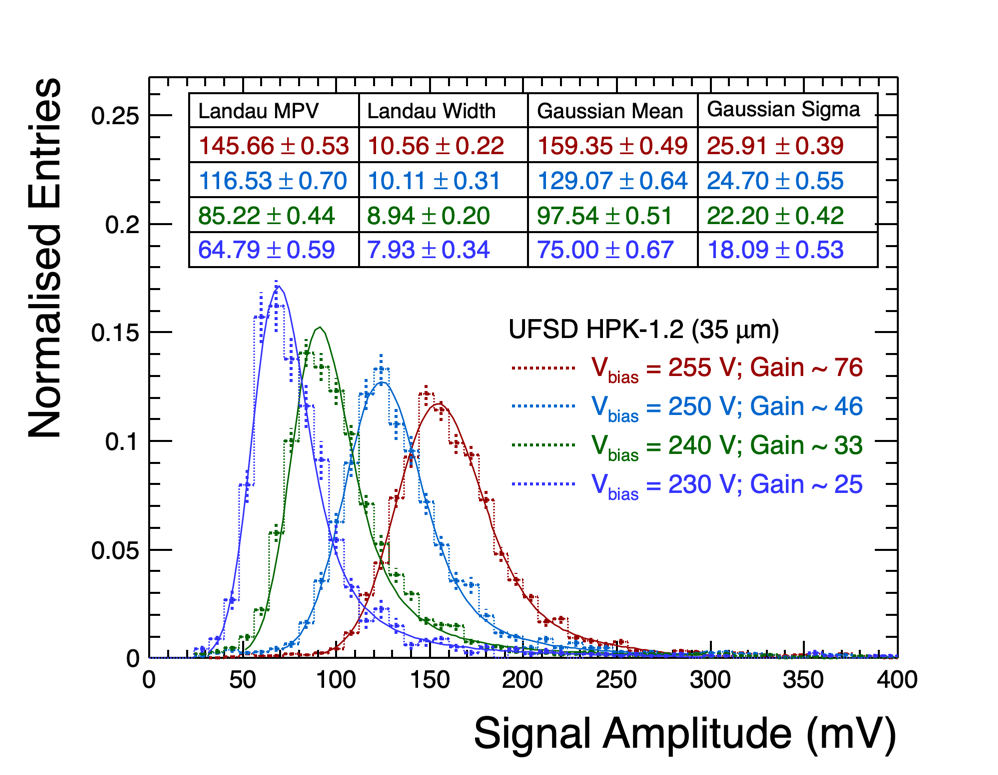}
\caption[Average signal waveform and normalized signal amplitude.]{Average signal shape (left) and distribution of signal amplitude normalized by its integral (right).}	
\label{fig:pulsesignalamplitude}
\end{figure}

\begin{figure}[ht!]
\centering
\includegraphics[width=0.5\textwidth,clip,trim=5mm 5mm 20mm 20mm]{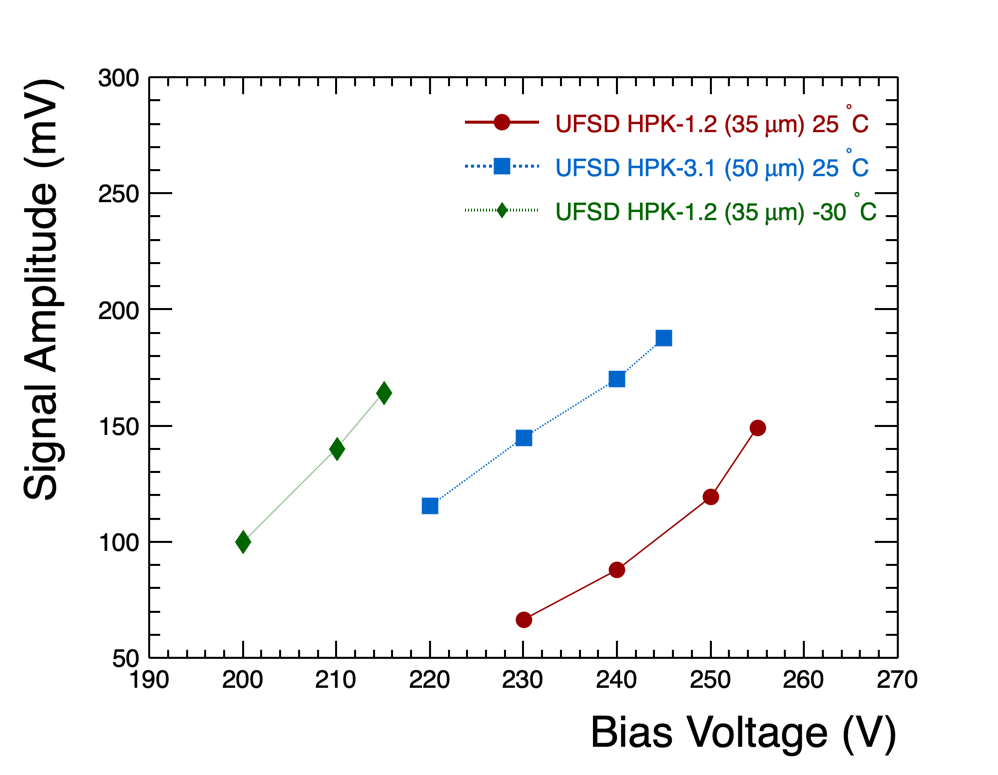}
\caption[Signal amplitude vs. Bias voltage.]{The pulse signal amplitude as a function of bias voltage.}
\label{fig:pulsesignalamplitudevsbiasvoltage}
\end{figure}

\nin In general, the energy deposited in the thin silicon sensor follows a Landau distribution. The average signal shape at different internal gains for HPK-1.2 at room temperature is shown in the left panel of figure~\ref{fig:pulsesignalamplitude}. 
However, the right plot in figure ~\ref{fig:pulsesignalamplitude} shows the convolution of Landau and Gaussian distributions of the signal amplitude for HPK-1.2 at room temperature. The studies performed with a laser pulse imply that the Gaussian contribution is from the read-out board's impulse response~\cite{Cartiglia:2016voy}. The Landau distribution provides the most probable value (MPV) and Full Width Half Maximum (FWHM). Whereas, Gaussian width, $\sigma_{Gauss}$ demonstrate the noise and fluctuations. The increase in the Gaussian width with bias voltage implies an increased contribution from the impulse response of the UFSD–amplifier board (shown in figure~\ref{fig:readoutboardfull}) and the sensor's shot noise. Figure~\ref{fig:pulsesignalamplitudevsbiasvoltage} shows the increasing trend of signal amplitude as a function of bias voltage for HPK-1.2 and HPK-3.1 at room temperature and -30 $^\circ$C. 

\subsection{Noise and Jitter}\label{subsec:noiseandjitter}

\begin{figure}[ht]
\centering
\includegraphics[width=0.5\textwidth,clip,trim=5mm 5mm 25mm 25mm]{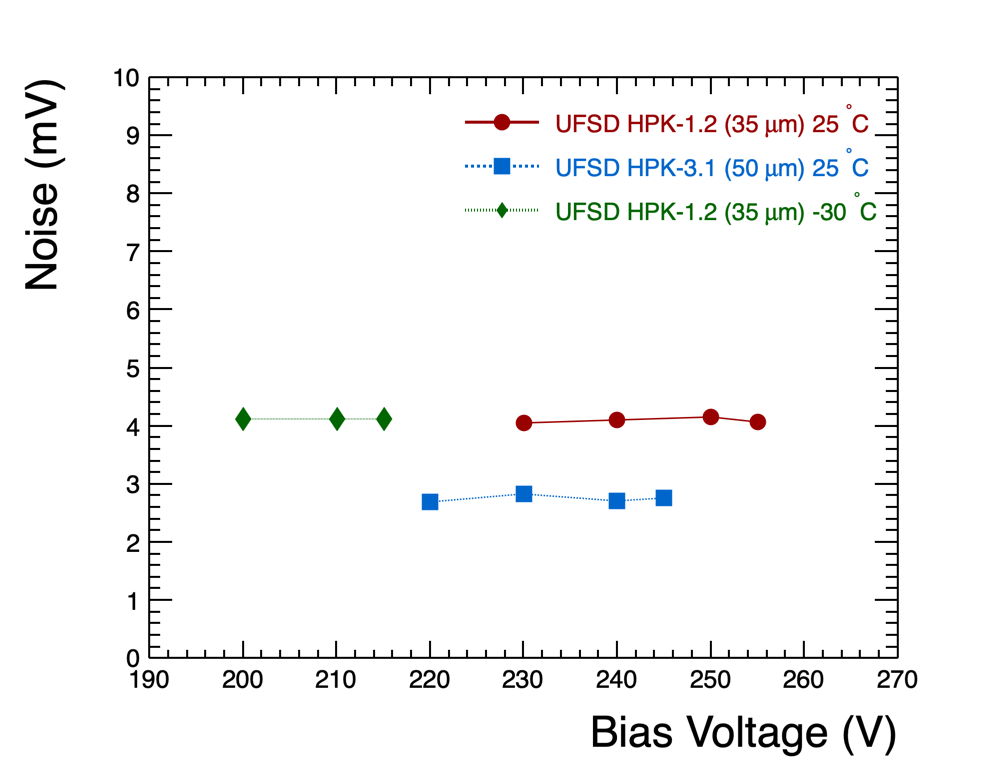}
\includegraphics[width=0.5\textwidth,clip,trim=5mm 5mm 25mm 25mm]{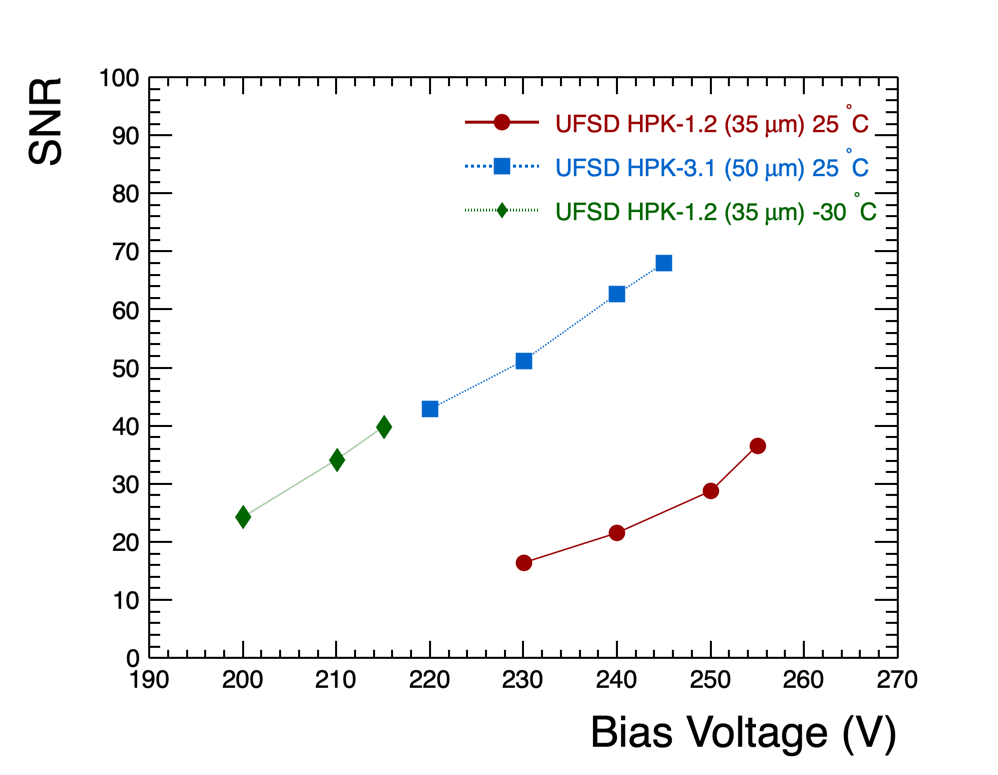}
\caption[Noise vs. Bias Voltage.]{The Noise (left) and Signal-to-Noise Ratio (right) as a function of bias voltage.}
\label{fig:rms}
\end{figure}

\nin The noise at different bias voltages is shown in the left panel of figure~\ref{fig:rms} for HPK-1.2 and HPK-3.1.
The RMS also includes the digitization noise from oscilloscope, the contributions of which varies depending on vertical scale (mV/div). It has been seen that the larger vertical scale produce larger noise~\cite{GALLOWAY201919}. The noise usually remains stable in the range of 2.5-4.5 mV throughout the beam test. The corresponding signal to noise ratio (SNR) increases with bias voltage and subsequently with the sensor's gain, as shown in the right panel of the figure~\ref{fig:rms}. The signal is given in the figure~\ref{fig:pulsesignalamplitudevsbiasvoltage}. It is also seen that the SNR at certain bias voltage increases at low temperature. The increase in SNR is mainly due to the increase in the gain at low temperature.   

\begin{figure}[ht!]
\centering
\includegraphics[width=0.5\textwidth,clip,trim=5mm 5mm 20mm 20mm]{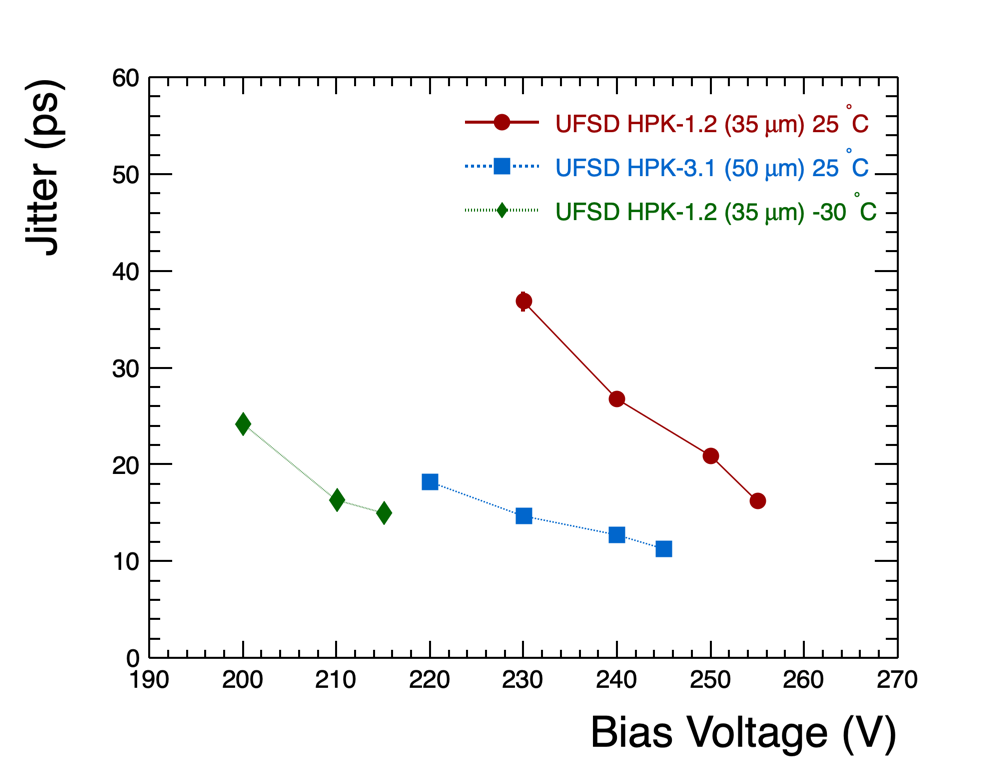}
\caption[Jitter vs. Bias Voltage.]{The Jitter as a function of bias voltage.}	
\label{fig:jitterbiasvoltage}
\end{figure}

Figure~\ref{fig:jitterbiasvoltage} shows the jitter as a function of bias voltage for HPK-1.2 and HPK-3.1 at room temperature and -30 $^\circ$C. It shows that the jitter decreases with increasing bias voltage. 

\subsection{Rise Time}\label{subsec:risetime}

\begin{figure}[!ht]
\centering
\includegraphics[width=0.5\textwidth,clip,trim=5mm 5mm 20mm 20mm]{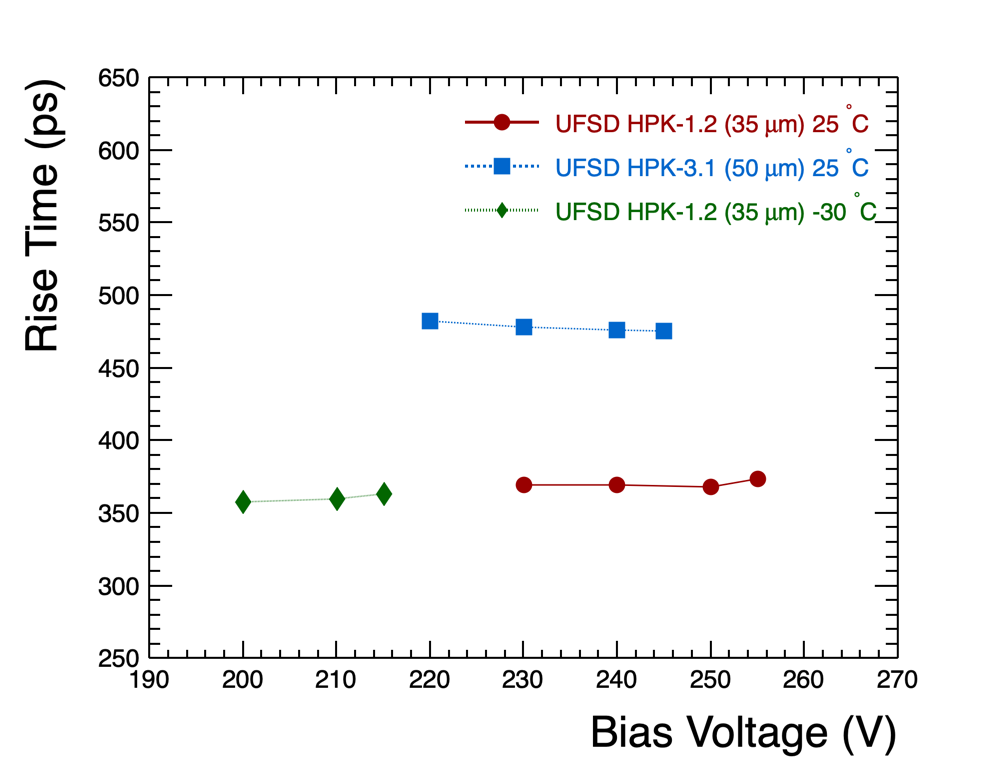}
\caption[Rise time vs. bias voltage.]{The rise time as a function of the bias voltage.}	
\label{fig:risetime}
\end{figure}

\nin The rise-time is a crucial parameter in defining the timing resolution of the UFSD, and it is determined by the drift time of the electrons. The rise-time is the time taken by the signal pulse to reach 90$\%$ of maximum signal amplitude from 10$\%$. Figure~\ref{fig:risetime} shows the rise-time at different bias voltages for HPK-1.2 and HPK-3.1 at room temperature and -30 $^\circ$C. The uncertainty over the measurement of rise-time is within 2 ps. The decreasing trend of the HPK-3.1's rise-time indicates an increase in the electric field in the bulk with bias voltage, due to which the hole drift velocity inside the bulk increases. A similar effect is observed while going from room temperature to low temperature measurements for HPK-1.2. Whereas, for HPK-1.2, the rise-time increases slightly, indicating delay due to the multiplication mechanism at higher gain. Figure~\ref{fig:risetime}, reflects these effects very minimally and the wide range of data in bias voltage is required to show it more prominently. 

\subsection{Timing Measurements}\label{subsec:timingmeasurements}

\begin{figure}[ht]
\centering
\includegraphics[width=0.5\textwidth,clip,trim=5mm 5mm 25mm 23mm]{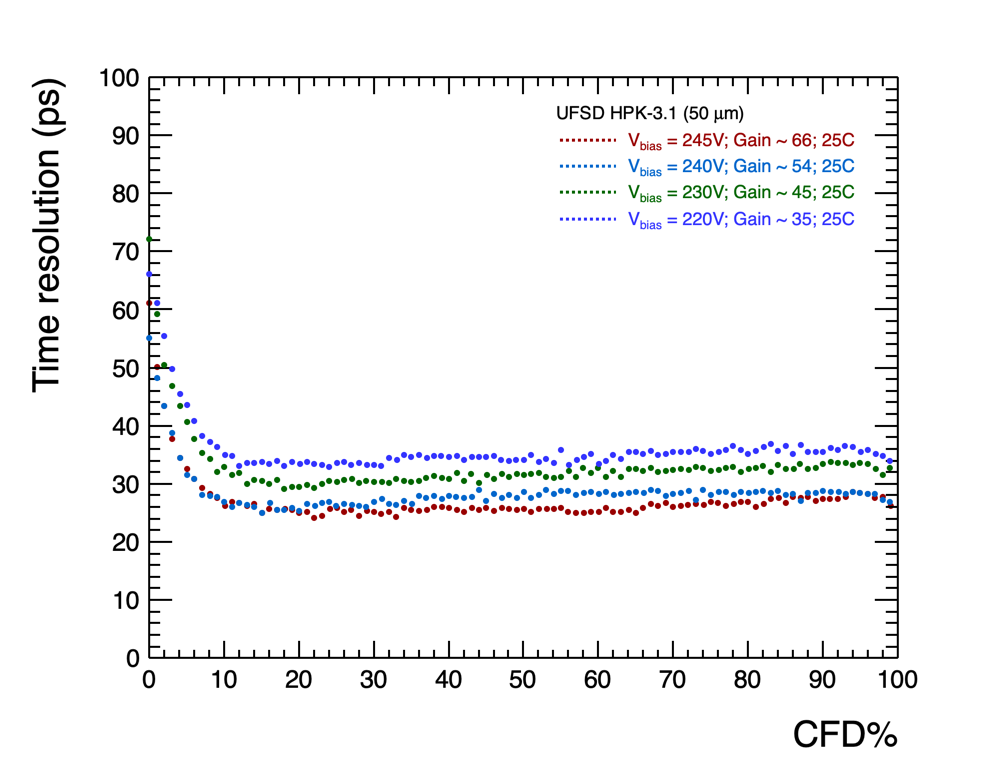}
\includegraphics[width=0.5\textwidth,clip,trim=5mm 5mm 25mm 23mm]{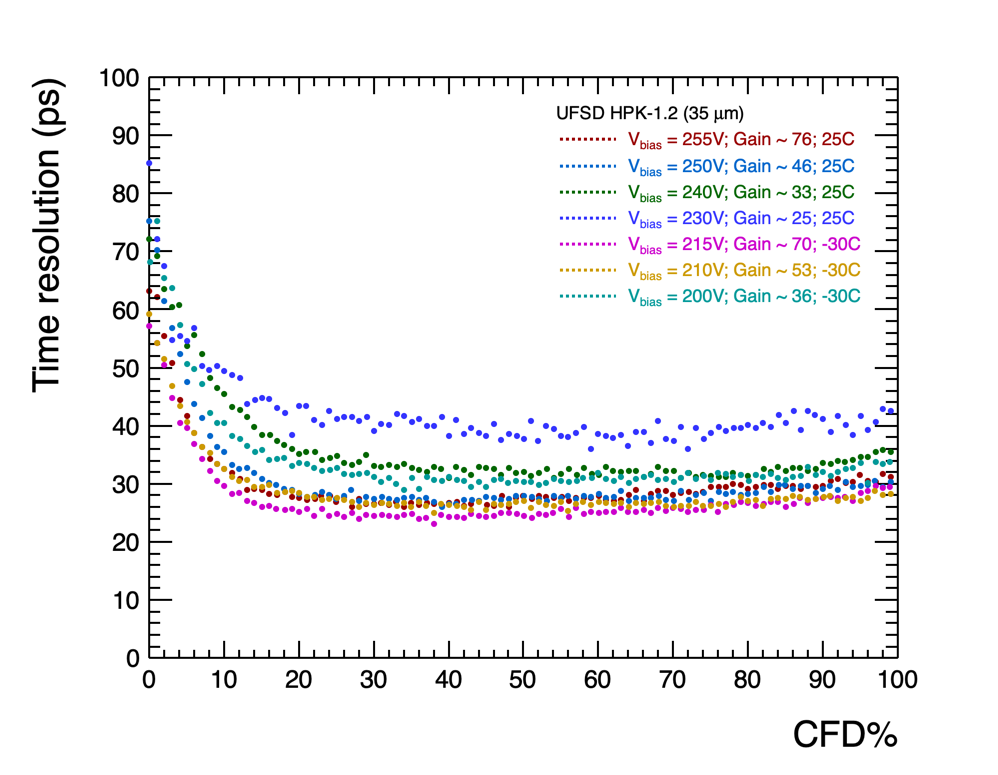}
\caption[Timing measurements vs. CFD.]{Timing measurement as a function of CFD fraction for HPK-3.1 at room temperature (left) and HPK-1.2 at room temperature and at -30 $^\circ$C (right).}
\label{fig:timevscfd}
\end{figure}

\begin{figure}[ht]
\centering
\includegraphics[width=0.5\textwidth,clip,trim=5mm 5mm 25mm 23mm]{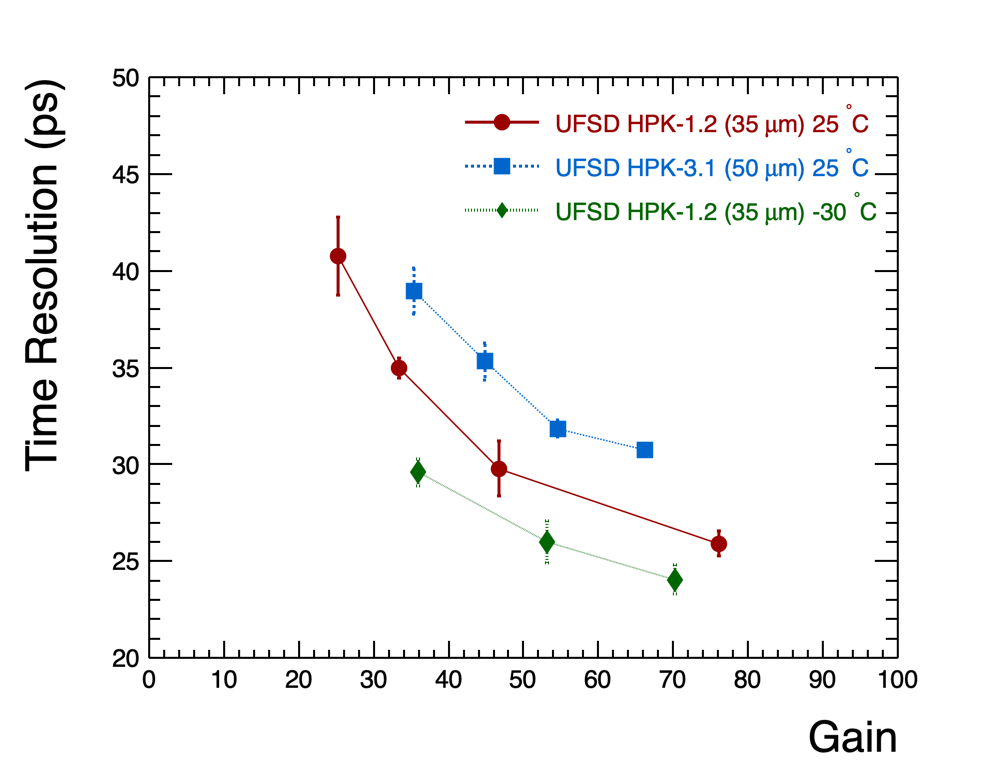}
\includegraphics[width=0.5\textwidth,clip,trim=5mm 5mm 25mm 23mm]{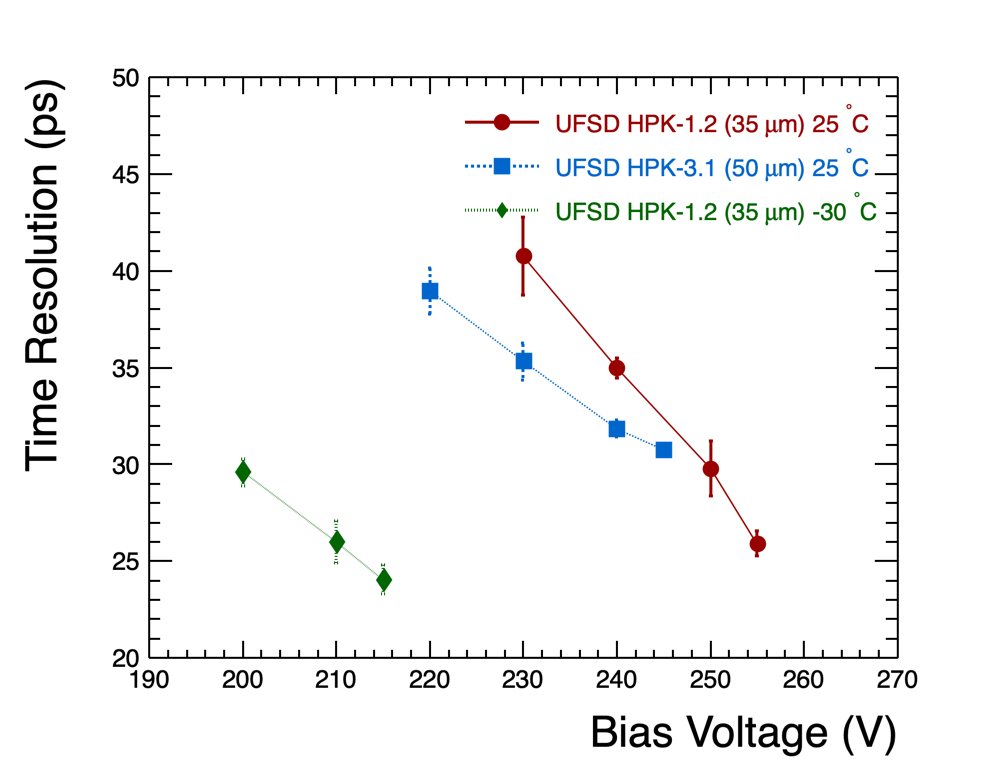}
\caption[Timing measurements for UFSDs and its temperature dependence.]{Timing measurement of UFSDs using time-of-flight (TOF) technique for minimum ionising particles (MIP). Timing resolution is shown as a function of gain (left) and bias voltage (right).}
\label{fig:timingandtemperature}
\end{figure}

\nin The timing resolution as a function of the CFD fraction is shown in figure~\ref{fig:timevscfd} for HPK-1.2 and HPK-3.1 at room temperature and -30 $^\circ$C. The measured timing resolution is stable above $\sim$15$\%$ up-to 80$\%$ of CFD fraction. The timing resolution of each of the LGADs described in section~\ref{subsec:lgaddevice} has been obtained using the CFD method and shown in figure~\ref{fig:timingandtemperature}. The results show timing performance at room temperature as well as at -30 $^\circ$C. 

In the beam test, 3 HPK-1.2 LGADs are used as DUTs, and a 4$^{th}$ LGAD (HPK-8664) is used as the trigger. The time resolution is calculated for ten sets of DUT combinations. The time differences from different combinations of DUTs and TRG are

\begin{description}
\item[$\Delta(t_{DUT}-t_{TRG})$:] Three sets of time difference between DUT and the trigger. 

These are the time differences calculated using single DUT, so called as "singlets".

\item[$\Delta(t_{\average{2{\rm DUT}}}-t_{TRG})$:] Three sets of time difference between average of pair of DUTs and the trigger. As the time difference is taken for average of two UFSDs they are called as "Doublets".
\item[$\Delta(t_{\average{3{\rm DUT}}}-t_{TRG})$:] The time difference between average of three DUTs and the trigger gives the "triplet" measurement.
\end{description}

\nin Along with this, the time resolution is calculated considering only DUTs as follows,
\begin{description}
\item[$\Delta(t_{DUT}-t_{DUT'})$:] Three sets of time difference between pairs of DUTs. 
\item[$\Delta(t_{\average{2{\rm DUT}}}-t_{DUT'})$:] Three sets of time difference between average of pair of DUTs and third DUT.
\end{description}

\begin{figure}[!ht]
\centering
\includegraphics[width=0.5\textwidth]{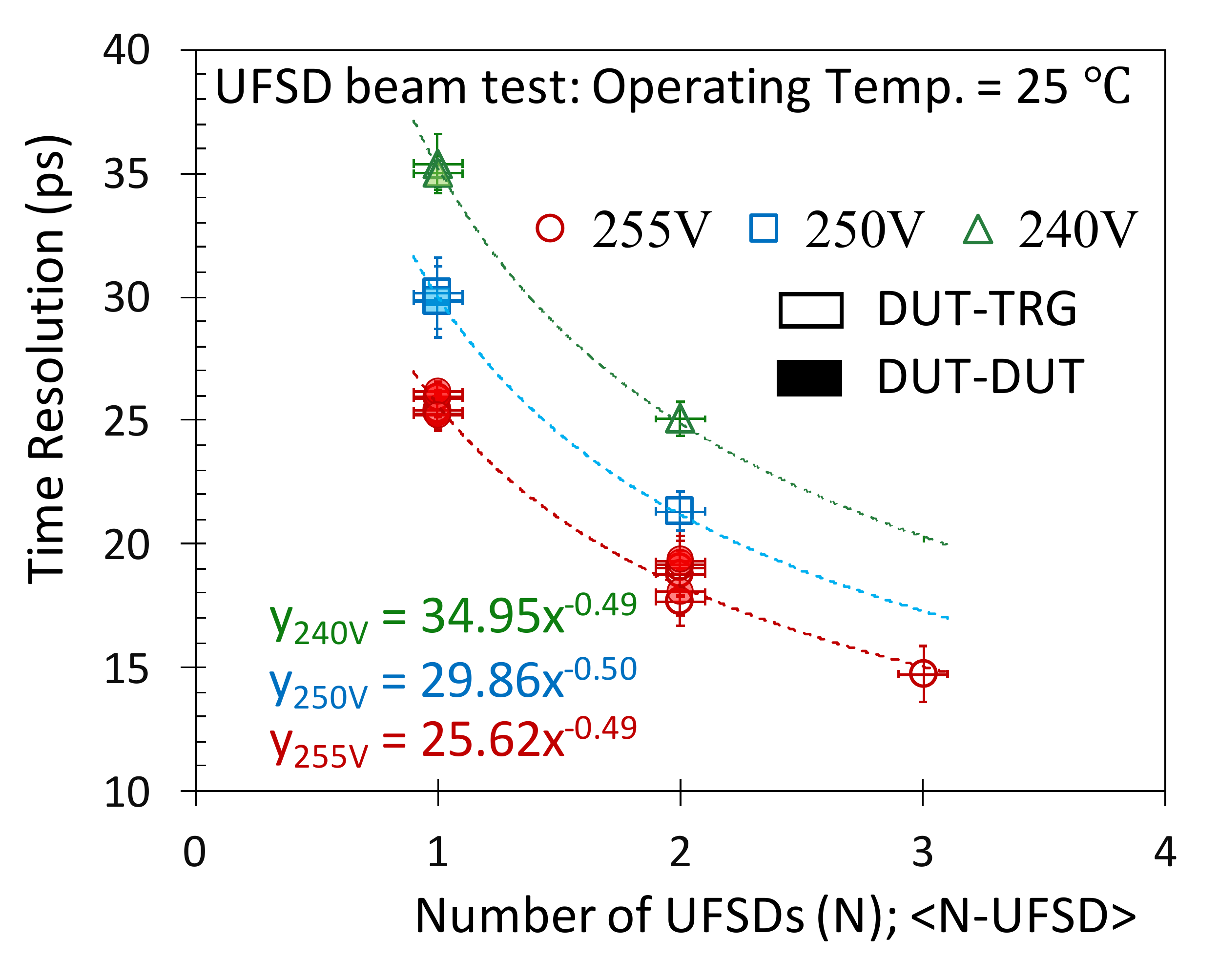}
\includegraphics[width=0.5\textwidth]{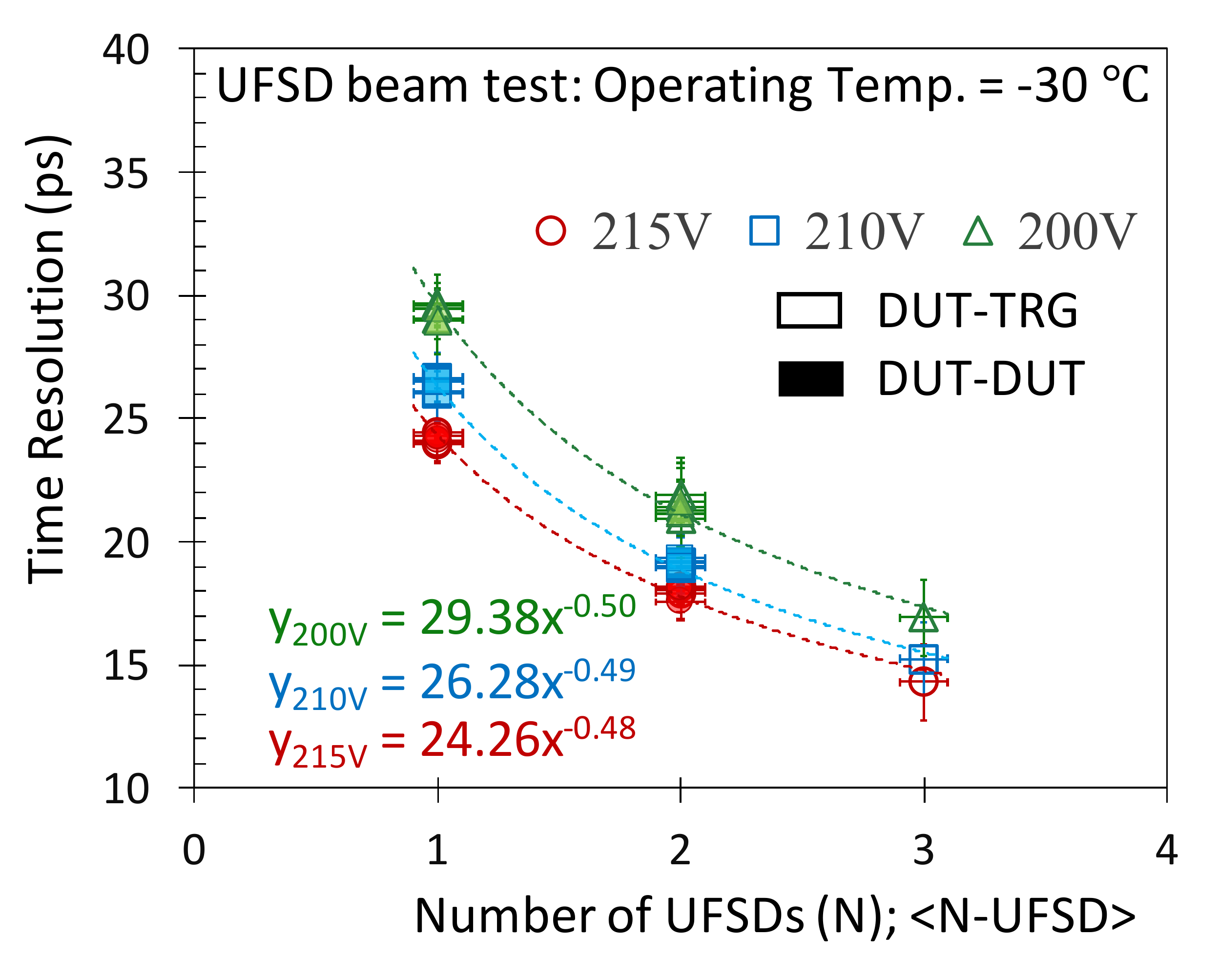}
\caption[Timing measurements for UFSDs and its temperature dependence.]{The timing resolutions at different bias voltages as a function of the number of UFSDs combined. The data reports measurements for the HPK-1.2 with thickness 35 \mum at room temperature (left) and -30 $^\circ$C (enclosure air temperature) (right) with a 120 GeV proton beam at the Fermilab Test Beam Facility. The missing data point (for V$_{bias}$ = 240 V and 250 V for N = 3 at 25 $^\circ$C) is due to the failure of power supply to DAQ during the test beam run.}
\label{fig:timeresolutioncombUFSD}
\end{figure}

The results for these measurements are shown in  figure~\ref{fig:timeresolutioncombUFSD} and tabulated in table~\ref{tab:timeresolution} for different operating conditions.
\begin{table}[!ht]
  \centering
  \caption[Timing resolution for set of DUTs]{The timing resolution for sets of UFSDs; a single (N=1), a pair (N=2), and the triplet (N=3). The data reports measurements for the HPK-1.2 with a 120 GeV proton beam at the Fermilab Test Beam Facility. The missing data point (for V$_{bias}$ = 240V and N = 3) is due to the failure of power supply to DAQ during the test beam run.}
  \label{tab:timeresolution}
  \begin{tabular*}{\textwidth}{@{\extracolsep{\fill}} cccc}
    \toprule
     \multirow{3}{*}{Number of DUTs}	&  \multicolumn{3}{c}{UFSD Timing Resolution (ps)} \\
      \cmidrule(lr){2-4}
  							& V$_{\rm bias}$ = 240 V & V$_{\rm bias}$ = 255 V & V$_{\rm bias}$ = 215 V \\
  							& (T = 25 $^\circ$C) & (T = 25 $^\circ$C) & (T = -30 $^\circ$C) \\
    \midrule
    \multirow{1}{*}{N = 1} 		       	& \multirow{1}{*}{35.1 $\pm$ 1.0} 	& \multirow{1}{*}{25.6 $\pm$ 0.5} 	& \multirow{1}{*}{24.2 $\pm$ 0.7} \\
     \multirow{1}{*}{N = 2} 			& \multirow{1}{*}{25.0 $\pm$ 0.7} 	& \multirow{1}{*}{18.7 $\pm$ 1.1} 	& \multirow{1}{*}{18.0 $\pm$ 0.9} \\
     \multirow{1}{*}{N = 3} 			& \multirow{1}{*}{-} 	& \multirow{1}{*}{14.7 $\pm$ 1.2} 	& \multirow{1}{*}{14.3 $\pm$ 1.5} \\
    \bottomrule
  \end{tabular*}
\end{table}  
The bias voltage close to the breakdown voltage is selected to get the highest possible gain and a comparison is carried out at two lower voltages. The measurements show results taken at room temperature and -30 $^\circ$C. The results are in good agreement with the basic expectation of scaling given by $\sigma({\rm N})=1/\sqrt{\rm N}$. At low temperatures, the saturation velocity of the electron is reached at the lower electric field compared to that of higher temperature. A better timing performance is observed by cooling, mainly because of the increase in the drift velocity and the gain. The higher drift velocity shortens the rise time, providing a faster response. The timing resolution of 14.3 $\pm$ 1.5 ps is measured from a triplet of 35 \mum thick UFSDs (HPK-1.2) operated with a bias voltage of 215 V at -30 $^{\circ}$C, which is the fastest reported timing resolution for UFSDs to date at a test beam.

\FloatBarrier

\section{Conclusions}\label{sec:conclusions}

\nin The UFSDs are tested for timing resolution with the 120 GeV proton beam at the Fermilab Test Beam Facility. The measurements are carried out for 35 \mum~(HPK-1.2) and 50 \mum~(HPK-3.1) LGAD sensors at room temperature and -30 $^\circ$C. It is seen that the thinner sensor provides a better timing response with increase in internal gain. The timing resolution also improves at low temperatures. The single LGAD sensor with 35 \mum thickness (HPK-1.2) and active area of 1.3$\times$1.3 mm$^2$ provided a timing resolution of 35.1 $\pm$ 1.0 ps and 25.6 $\pm$ 0.5 ps at 240 V and 255~V respectively at room temperature. The same LGAD sensor's timing performance is improved to 24.2 $\pm$ 0.7 ps when operated at -30 $^{\circ}$C. In the test beam, a telescope comprising four UFSD planes (3 DUTs and 1 Trigger) is tested demonstrating the improvement of the timing resolution given by the relation 1/$\sqrt{N}$, where N is the number of UFSDs. When three planes of HPK-1.2 LGADs are averaged (triplet), the timing performance of 14.3 $\pm$ 1.5 ps is measured with a reverse bias voltage of 215~V and at a temperature of -30 $^\circ$C, which is the fastest reported timing performance to date. It is observed that the targeted timing response of 10 ps is achievable with an optimized thickness and/or number of layers of LGAD sensors.
\FloatBarrier

\section{Acknowledgements}\label{sec:acknowledgements}

\nin This work is supported by Laboratory Directed Research
and Development (LDRD) funding, "Tomography at an Electron-Ion Collider: Unraveling the Origin of Mass and Spin" from Argonne National Laboratory, provided by the Director,
Office of Science, of the U.S. Department of Energy under Contract No. DE-AC02-06CH11357. This work is also supported by the United States Department of Energy, grant DE-FG02-04ER41286.

This document is prepared  using the resources of the Fermi National Accelerator Laboratory (Fermilab), a U.S. Department of Energy, Office of Science, Fermi Lab Test Beam Facility. Fermilab is managed by Fermi Research Alliance, LLC (FRA), acting under Contract No. DE-AC02-07CH11359.

We are thankful to Thomas O'Connor and his team at the Physics Division, Argonne National Laboratory for their help in fabricating the alignment box. We also like to acknowledge the help provided by Michelle Jonas and Humberto Gonzalez at the Silicon Detector Facility at Fermilab in wire-bonding LGAD sensors onto the read-out boards. 
\FloatBarrier

\bibliography{mybibfile}

\end{document}